\documentclass[aps,prd,twocolumn,showpacs,amsmath,amssymb,nofootinbib]{revtex4}
\usepackage[utf8]{inputenc}
\usepackage[english]{babel}
\usepackage[dvips]{graphicx}
\usepackage{empheq}

\newcommand{\om} \omega   \newcommand{\Om} \Omega
\newcommand{\eps} \epsilon

\newcommand{\be} {\begin{equation}}
\newcommand{\ee} {\end{equation}}
 \newcommand{\ba} {\begin{eqnarray}}
 \newcommand{\ea} {\end{eqnarray}}
\newcommand{\bal} {\begin{align}}
\newcommand{\eal} {\end{align}}

\def\lrD{\mathrel{{\cal D}\kern-1.em\raise1.75ex\hbox{$\leftrightarrow$}}}
\def\lr #1{\mathrel{#1\kern-1.25em\raise1.75ex\hbox{$\leftrightarrow$}}}
\newcommand{\eqr}[1]{Eq.~(\ref{#1})}
\newcommand{\figr}[1]{Fig.~\ref{#1}}

\newcounter{footref}

\begin{document}
\title{Black/White hole radiation from dispersive theories}

\author{Jean Macher}
\email{jean.macher@th.u-psud.fr}
\affiliation{Laboratoire de Physique Th\'eorique, CNRS UMR 8627, B\^at. 210, Universit\'e Paris-Sud 11, 91405 Orsay Cedex, France}
\author{Renaud Parentani}
\email{renaud.parentani@th.u-psud.fr}
\affiliation{Laboratoire de Physique Th\'eorique, CNRS UMR 8627, B\^at. 210, Universit\'e Paris-Sud 11, 91405 Orsay Cedex, France}

\date{\today}

\begin{abstract}
We study the fluxes emitted by black holes when 
using dispersive field theories. 
We work with stationary one dimensional backgrounds which are 
asymptotically flat on both sides of the horizon. 
The asymptotic fluxes are 
governed by a $3\times3$ Bogoliubov transformation. 
The fluxes emitted by the corresponding white holes 
are regular and governed by the inverse transformation. 
We numerically compute the spectral 
properties of these fluxes for both sub- and superluminal quartic dispersion. 
The leading deviations with respect to the dispersionless flux 
are computed and shown to be governed by a critical frequency 
above which there is no radiation. Unlike the UV scale governing dispersion, 
its value critically depends on the asymptotic properties of the background. 
We also study the flux outside the robust regime. In particular
we show that its  low-frequency part remains almost thermal but with a 
temperature which significantly differs from the standard one. 
Application to four dimensional black holes and Bose Einstein 
condensates are in preparation. 
\end{abstract}

%04.70.Dy=Quantum aspects of black holes, evaporation, thermodynamics
%04.60.Bc 	Phenomenology of quantum gravity %J5 pour les mini-BH au LHC
%43.35.+d=Ultrasonics, quantum acoustics, and physical effects of sound
\maketitle

\section{Introduction}

In 1981, Unruh showed~\cite{Unruh:1980cg} that the phonon field equation in a 
nonhomogeneous flow is analogous to that of a relativistic field 
propagating in a four dimensional curved space-time. He also suggested 
that this analogy could be used to experiment black hole (BH) radiation in the lab. 
However, as pointed out in \cite{Jacobson:1991bh}, 
this analogy is limited by the fact that in condensed matter the phonon field 
becomes dispersive above a certain wave vector, say $\Lambda$. 
Motivated by this remark, Unruh numerically showed that the fluxes 
emitted by an acoustic black hole are not significantly affected 
by the dispersive properties of the phonons, 
at least when $\Lambda \gg \kappa$, 
where $\kappa$ is the gradient of the fluid velocity at the sonic horizon.~\cite{Unruh:1994je} 
Since then, this robustness of the fluxes has been confirmed both by algebraic~\cite{Brout:1995wp,Corley:1997pr,Himemoto:1999kd,Saida:1999ap,Unruh:2004zk,Balbinot:2006ua} 
 and numerical~\cite{Corley:1996ar,Unruh:2007zz}  techniques. 

However, many open questions remain. In particular, little is known about 
\begin{itemize}

\item the exact nature and the precise range of the 
parameters delimiting the robust regime,

\item the scaling properties of the leading deviations in the robust regime,

\item the properties of the fluxes outside this regime, and

\item the comparison of sub- and superluminal dispersion.

\end{itemize}
In view of the 
possibility to detect Hawking radiation 
in the lab,~\cite{Barcelo:2005fc} 
it is important to provide quantitative estimates
of the modifications of the fluxes 
induced by dispersion. Besides this reason, answering the above questions is also 
relevant when exploring the consequences on black hole physics due to violations of 
Lorentz invariance~\cite{Jacobson:1991bh,Jacobson:2005bg,Parentani:2000ts,Parentani:2007mb,Horava:2009uw} 
that could stem from (or be related to) quantum gravity. 
This could be true in particular when considering 
the signals emitted during the evaporation of 
light black holes ($M\sim 10^{-16}$ Planck mass) 
that could be produced at the LHC~\cite{Yoshino:2005hi}.

In this paper, we aim to answer these questions by a two step analysis. 
First we provide a complete description of 
the quantum propagation of a dispersive field in stationary black hole geometries. 
(To our knowledge, this has not been presented anywhere.)
Secondly, focussing on quartic dispersion and
using numerical methods, we study both the small modifications 
of the fluxes induced by dispersion when $\Lambda \gg \kappa$, 
and the larger modifications when this inequality is not satisfied. 

We are planning to extend this 
work in two directions. First, to Bose condensates, where the phonon field obeys the Bogoliubov-de-Gennes equation 
which differs from the ``scalar'' field 
equation we consider below (this program has been completed during the publication process of the present work~\cite{MacherParentaniBEC}) and  secondly to 4D black holes, where the gravitational potential  
induces grey body factors which could play a critical role. 

In Sec.~\ref{settings}, we present the class of velocity profiles and the stationary mode equation.
In Sec.~\ref{analytic}, we study the algebraic properties of the modes and the 
Bogoliubov transformation one is dealing with. In Sec.~\ref{numproc}, we present our numerical treatment.
In Secs.~\ref{numressub} and~\ref{numressuper}, we numerically solve the mode equation
and interpret the results for sub- and superluminal quartic dispersion respectively. 
We conclude in Sec.~\ref{conclusions}.

\section{The settings\label{settings}}

\subsection{Velocity profiles}

We work in two space-time dimensions and consider stationary black hole 
geometries. These shall be characterized by the velocity profile $v(x)$:
\be
ds^2 = -dt^2 + (dx - v(x)dt)^2.\label{metric}
\ee
This expression appears~\cite{Unruh:1980cg} when considering the 
propagation of low-frequency phonons in a moving fluid whose 
velocity field $v(x)$ is measured in the 
Galilean frame (not necessarily the lab frame) where $v$ only depends on $x$.
The  sound velocity is assumed to be constant and has been set to $1$. 
A more general case will be considered in~\cite{MacherParentaniBEC}.
The above metric possesses an event horizon where the flow becomes supersonic.

When the fluid flows to the left,
$v < 0$, and when $|v|$ increases toward the left,
one obtains a future (black hole BH) horizon.
Had we considered the 
flow $-v(x)$, we would have obtained a white hole (WH) geometry.
As we shall see, the spectral properties (in terms of the conserved frequency 
$\om = i \partial_t$) 
of fluxes emitted by this WH are in one-to-one correspondence with
those of the fluxes emitted by the BH governed by $v(x)$. However they differ slightly. 
It should also be pointed out that, since we only consider dispersive mode equations, 
see e.g. \eqr{weq0}, $-\infty < t, x < \infty$ in  \eqr{metric} 
represents the whole space-time~\cite{Jacobson:1998he}, 
and not only a part of it as would have been the case when using relativistic fields.

We shall also assume that $v$ becomes asymptotically constant on both sides of the horizon. 
The existence of two asymptotic regions is important 
because one then has well-defined asymptotic modes on {\it both sides},
and therefore well-defined particle fluxes.
This is particularly relevant for the proposal~\cite{Balbinot:2007de,Carusotto:2008ep} 
based on measurements of the long distance correlations~\cite{Massar:1996tx,Brout:1995rd} 
between Hawking quanta and their partners propagating
on the other side of the horizon.

In order to determine the
\emph{generic} consequences of dispersion, we shall consider a large class 
of flows.
In contrast with the nondispersive case where they play no role,
we shall see that the asymptotic properties of $v(x)$
are crucial when using nonlinear dispersion relations.
This will imply in particular that the inequality  $\Lambda \gg \kappa $ is not sufficient to guarantee no significant deviation from the standard fluxes.

\begin{figure}
\includegraphics{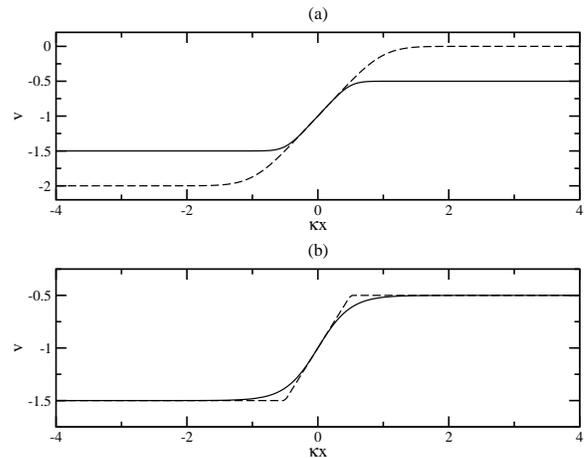}
\caption{Velocity profile $v(x; \kappa, n, D)$ as a function of $\kappa x$. 
(a) $n$ fixed to 2, $D=0.5$ (solid line) and $D=1$ (dashed line). (b) $D$ fixed to $0.5$, $n=1$ (solid line), $n=20$ (dashed line).\label{fig::vd}}
\end{figure}

The class we shall use generalizes that of 
\cite{Corley:1996ar} and contains profiles of the form
\be
v(x) = -1 + D \,
{\rm sign}(x) \, \tanh^{1/n}\left[\left(
\frac{\kappa |x| }{D}\right)^n\right].\label{vdparam}
\ee
When changing $n$ and $D$, $\kappa$, the slope of $v$ at the horizon, is fixed,
as is the location of the horizon at $x=0$.
When using the 2D massless relativistic theory, the emitted spectrum 
only depends on $\kappa$. Instead, when using dispersive 
theories, together with the ratio $\Lambda/\kappa$, both $n$ and $D$
also affect the properties of the fluxes. 
Figure~\ref{fig::vd} shows $v(x)$ for several values of $D$ and $n$, as a function of $\kappa x$. 

The power $n$ controls the sharpness of the transition from the linear behavior of $v$ near the horizon 
to the asymptotic flat region. For $n \to \infty$, the transition becomes sharp. 
When using dispersive theories, 
sharp transitions give rise to nonadiabatic effects which produce superimposed 
oscillations~\cite{Corley:1996ar}.
Unless said otherwise, $n$ is equal to $2$ in the following.

The parameter $D\in \, ]0,1]$ fixes the asymptotic values of the velocities,
\be
v_\pm = -1 \pm D,
\ee
for $x \to \pm \infty$ respectively.
When using nonlinear dispersion relations,
$D$ fixes the critical frequency $\om_{\rm max}$
above which the radiation 
vanishes. This is relevant because $D$ is constrained to be much smaller than $1$ 
in some proposals of experiments~\cite{Philbin:2007ji}. More generally, we shall find that $\om_{\rm max}$ is the most 
relevant parameter, in that the corrections induced by dispersion are essentially
governed by $\kappa/\om_{\rm max}$,
and not by $\kappa/\Lambda$ as one might have expected.

\subsection{Dispersive wave equation}

In Nature of course,
various dispersion relations are found depending on 
the condensed material used.
However, when imposing analyticity in $k^2$,
quartic dispersion relations represent the first two possible deviations from the linear case. 
(Dissipative theories constitute another class
that should be considered separately~\cite{Parentani:2007dw,Adamek:2008mp}.)
In this paper, we shall thus restrict our numerical analysis to 
\be
\Om^2 = k^2 \pm \frac{k^4}{\Lambda^2},\label{quartic}
\ee
where $\Om$ and $k$ are the energy and momentum measured in the frame comoving with the flow. 

When working with the metric of \eqr{metric} the corresponding wave equation for the 
velocity potential $\phi$ reads~\cite{Unruh:1994je} 
\ba
(\partial_t + \partial_x v)(\partial_t + v\partial_x)\phi = (\partial_x^2 \mp \frac{1}{\Lambda^2}\partial^4_x )\phi.
\label{weq0}
\ea
when assuming that the fluid density is constant. (This restriction will also be removed
in~\cite{MacherParentaniBEC}.)
The upper (lower) sign corresponds to super- (sub-) luminal dispersion. 
The conserved scalar product takes the form 
\be
(\phi_1,\phi_2) = i\int_{-\infty}^{\infty}dx \left[\phi_1^*(\partial_t + v\partial_x)\phi_2 - \phi_2(\partial_t + v\partial_x)\phi_1^*\right].\label{KGnorm}
\ee
It is independent of the dispersion and it coincides with the standard
Klein-Gordon (KG) product evaluated on the preferred time slices specified by dispersion (here $t= cst$).

Since the flow is stationary, we shall work at fixed (Killing) frequency $\om$. 
Setting $\phi = e^{-i\om t}\phi_{\om}(x)$ in the wave equation 
yields
\ba
(-i \om + \partial_x v)( -i \om + v \partial_x)\phi_\om =
(\partial_x^2 \mp \frac{1}{\Lambda^2}\partial^4_x )\phi_\om.
\label{waveeq}
\ea
As usual, the scalar product is block diagonal in $\om$. 
Therefore, when working with stationary states, all observables 
will be expressed as sums 
of observables defined at fixed $\om$.
More precisely, one should consider the couple of frequencies
$(\om, -\om)$ since opposite values of $\om$ mix in stationary Bogoliubov transformations. 
In what follows, to avoid any confusion, $\om$ is always positive definite.

Before numerically studying the solutions of Eq.~(\ref{waveeq}), it is worth 
analyzing 
the enlarged set of solutions 
and the Bogoliubov transformation
 one obtains.

We remind the reader that the oscillatory solutions to \eqr{waveeq} can be classified 
according to the sign of their group velocity in the comoving frame, $v_{gr}^{com}=d\Om/dk$. In the following we call the modes with $v_{gr}^{com}<0$ left movers and those with the opposite sign right movers. 
In the absence of dispersion (and for 2D massless fields), these two sectors are decoupled, and the Bogoliubov transformation is particularly simple. When introducing (quartic) dispersion,  four independent solutions to 
\eqr{waveeq} exist. 
The classification between left and right movers can still be done, but the left-right decoupling is
lost. It is worth mentioning that, when considering a varying speed of sound rather than a varying flow-velocity, 
see e.g.~\cite{Carusotto:2008ep}, this decoupling is lost even in the long wavelength approximation ({\it i.e.} without dispersion). 
Notice also that 
one can introduce dispersion in a way that preserves this factorization~\cite{Brout:1995wp,Schutzhold:2008tx}, 
but these models do not seem to govern condensed matter systems.

\section{Analytic study\label{analytic}}

In several aspects, the material presented below closely follows 
what has been presented in~\cite{Corley:1996ar}. The novelties are related to the fact that
we treated both asymptotic regions on an equal footing. In particular,
we establish the complete character of the ``in'' and ``out'' mode bases,
and we write the unitary Bogoliubov transformation which
governs the general case (for stationary, one dimensional, single horizon geometries).
Finally, we show how to relate the fluxes emitted by black and white holes.

\subsection{Asymptotic solutions}

Given our velocity profile, $v$ is asymptotically constant
in both regions $|\kappa x| \gg D$.
There,
the 
solutions of Eq.~(\ref{waveeq})
are superpositions of plane waves $e^{i k x}$  with constant amplitudes. 
To characterize a solution, one needs 
the roots $k(\om)$
of the asymptotic wave vectors and the amplitudes $A_k$ of the corresponding waves. 
Owing to the nonlinearity 
of the dispersion relation, in addition to the 
oscillatory modes, exponentially growing/decaying modes exist, governed by complex values of $k$.

We first study how to handle these extra modes in the present settings,
paying attention to the completeness of the mode basis.
To be specific, we concentrate on superluminal cases, 
and only indicate the (small) differences that arise in subluminal cases.
\begin{figure}
\includegraphics[scale=0.8]{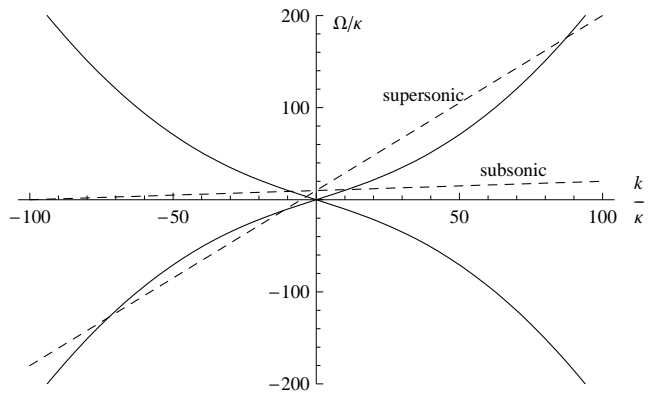}
\caption{\label{fig::asymptsols} 
The two straight lines represent $\om-v_+k$ for the subsonic one and $\om-v_- k$ for the supersonic one. The roots of \eqr{reldisp} correspond to the abscissa of the intersections of these lines with the curves $\pm\Om(k)$.
The numerical values are $p=1$ (quartic dispersion),
$\Lambda/\kappa=50$, $D=0.9$, $\om/\kappa=10$.}
\end{figure}
Let us thus analyze the roots of 
\be
(\om - v_{\pm} k )^2 
= k^2 +  \frac{k^{2p + 2}}{ \Lambda^{2p }} = \Omega^2(k).
\label{reldisp}
\ee
for $\om > 0$, and for an arbitrary value of the integer $p > 0$.
We work with a polynomial form so as to know the number of complex roots of \eqr{reldisp}. For everything concerning the real roots, 
nothing changes if one uses any strictly convex function $\Om^2(k)$ with a slope equal to 1 for low $k$.

In the subsonic region, $\vert v_+\vert < 1$, as can be seen in \figr{fig::asymptsols}, 
two real roots exist:  
$k^u_\om > 0$ and $k^v_\om <  0$ which correspond to a right and 
a left mover respectively. 
(Whenever an ambiguity could exist, we shall add a superscript $u$ or $v$ to designate 
right or left movers.)
$p$ pairs of complex conjugated roots also exist, since the equation is real.
One thus has $p$ roots with a positive imaginary 
part (which correspond to
growing modes to the right)
and $p$ solutions with a negative imaginary 
part (decaying modes). 

In the supersonic region $\vert v_-\vert > 1$, for $\om$ 
smaller than a critical frequency $\om_{\rm max}$ 
(we shall compute its value below)
there exist four real roots (see \figr{fig::asymptsols}), and thus only $p-1$ pairs of complex solutions.
In other words, when comparing this 
with what prevailed in the subsonic region, 
a pair of growing and decaying modes has been replaced
by a couple of oscillatory modes. Such replacements of pairs of modes
is a generic feature of QFT in external fields, 
see the Appendix of~\cite{Fulling:1989nb}. 
In the present hydrodynamical settings it will occur at all
horizons (for both black and white holes). 
For flows to the left, $v < 0$, the two new real roots both correspond to right movers.
Notice that for subluminal dispersion, 
the replacement occurs the other way around since the four real roots 
are found where the flow is subsonic.
 
Now, given that there are $2p + 2$ roots,
the 
general solution $\phi_\om(x)$ is 
a superposition of $2p + 2$ modes. 
However, when considering the Fourier transform
of the field operator, $\hat \phi_\om(x)$, only a subclass of these should  be used. 
And this subclass is \emph{complete} in a precise sense we explain 
below. 
(To our knowledge, this crucial aspect has not been discussed 
in the context of acoustic black holes.)
To establish the completeness of the mode basis
we first need to compute the frequency $\om_{\rm max}$
which is introduced by both sub- and superluminal dispersion, and which
will cut off Hawking radiation.

\subsection{Maximal frequency\label{ommexistence}}

\begin{figure}
\includegraphics[scale=0.8]{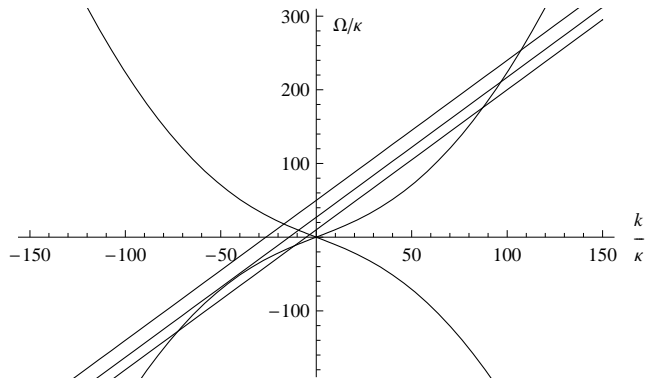}
\caption{\label{fig::merging}Solutions to \eqr{reldisp} in the supersonic region, for different values of $\om$. For $\om<\om_{\rm max}/\kappa$, 
two roots exist in the bottom left quadrant.
They move closer to one another as $\om$ increases, and merge when the straight line $\om - kv_-$ becomes tangent to $-\Om(k)$, for $\om=\om_{\rm max}$. For $\om>\om_{\rm max}$, these roots no longer exist.}
\end{figure}

The frequency $\om_{\rm max}$ is the value of $\om$ where the two extra real roots merge into
each other.
It is thus reached when the straight line $\om - v_{-}\, k$ is tangent 
to $-\Om(k)$,  as illustrated in \figr{fig::merging},
or equally 
when 
$-\vert \om\vert - v_{-}\, k$ is tangent to $\Om$,
where $\Om$ is the square root of the rhs of \eqr{reldisp}.

Restricting attention to quartic dispersion, 
the corresponding value of $k$ is 
\be
k_{\rm max, +} = \frac{\Lambda}{2\sqrt 2}\sqrt{v_{-}^2 - 4 + |v_{-}|\sqrt{v_{-}^2+8}}\, .
\ee
The frequency $\om_{\rm max}$ is then obtained from
\eqr{reldisp}.
It is thus of the form
\be
\om_{\rm max, +} = \Lambda f_{\rm +}(D).
\ee
For $D\ll 1$, one has $f_{\rm +}(D)\propto D^{3/2}$. 
This means that, whatever the value of the dispersion scale $\Lambda$, 
$\om_{\rm max, +}$ can be arbitrarily small, 
and in particular it could be even smaller than the 
temperature $\kappa/2 \pi$ one would get in a dispersionless theory.

For subluminal dispersion, the maximal frequency $\om_{\rm max, -}$
is again defined by the merging of the two extra real roots. In this case it occurs 
where 
$|v_+ | < 1$. 
In the quartic case the 
corresponding value of $k$ is
\be
k_{\rm max, -} = \frac{\Lambda}{2\sqrt 2}\sqrt{4-v_{+}^2 - |v_{+}|\sqrt{v_{+}^2+8}}.
\ee
Thus $\om_{\rm max, -}$ is also of the form $\om_{\rm max, -}
=\Lambda f_{\rm -}(D)$.
For $D \ll 1$, we 
again obtain $f_-(D)\propto D^{3/2}$.

\begin{figure}
\includegraphics[scale=0.8]{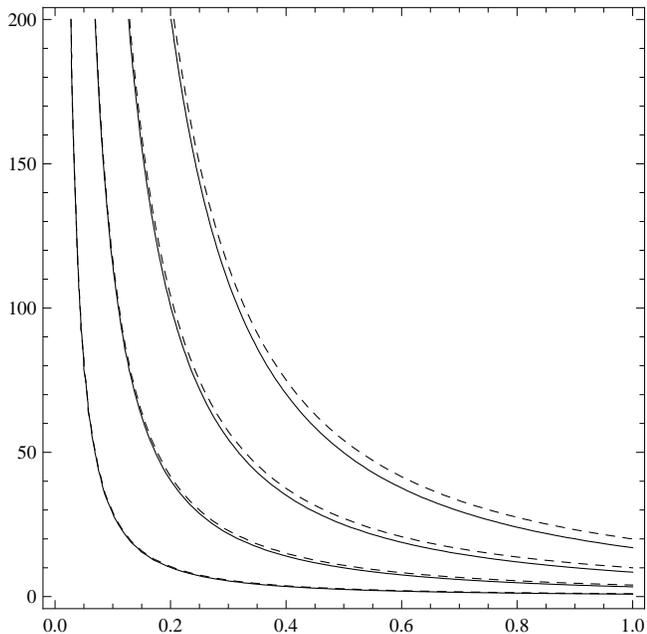}
\caption{\label{fig::ommaxcontours}Contours of constant $\om_{\rm max}$ in the plane $(D,\Lambda/\kappa)$. $D$ is the horizontal coordinate, $\Lambda/\kappa$ the vertical one. From left to right, the contours correspond to $\om_{\rm max}/\kappa=0.5,2,5,10$. The solid contours are for superluminal dispersion and the dashed ones for subluminal dispersion. The contours coincide for small $D$, and so do 
the functions $f_+$ and $f_-$. Their difference becomes visible as $D$ grows.}
\end{figure}

\figr{fig::ommaxcontours} shows the contours of constant $\om_{\rm max}$ in the plane $(D,\Lambda)$, 
for sub- and superluminal dispersion. The same value of $\om_{\rm max}/\kappa$ can be obtained with very different combinations of $(D,\Lambda/\kappa)$. 
As we shall see in Sec.~\ref{numressub}, the fluxes 
are highly degenerate on 
these contours, which means that the corrections with respect to the standard fluxes
are essentially governed by $\om_{\rm max}/\kappa$.

\subsection{Mode orthonormality and mode completeness\label{modecompleteness}}

To proceed to the canonical quantization, 
one should use mode bases that are 
orthonormal and complete. 
When the background
geometry is homogeneous, 
this is rather easy, even in the presence of dispersion. 
However, when $v$ of \eqr{metric} varies and in particular when it characterizes a horizon,
it becomes trickier.
Let us thus first consider situations where
$v$ is constant
both in space and time.

\subsubsection{Homogeneous metrics, $k$-representation}

In this case one should exploit the homogeneity and express
 the field operator in terms of exponentials $e^{ikx}$ and creation/destruction operators labelled by (real)
$k$ 
\be
\hat \phi(t,x) = \int_{-\infty}^\infty \frac{dk}{\sqrt{2 \pi}}  \left[  \hat a_k  \, e^{i k x} \, 
\frac{e^{- i \om t}}{\sqrt{2 \Om(k)}} + h.c. \right],
\label{fop}
\ee
where $ \om = \Om(k) + v  k$ with $\Om(k)$ given in Eq.~(\ref{reldisp}).
In this representation, creation/destruction operators obey the usual 
commutators
$[\hat a_k, \hat a^{\dagger}_{k'}] = \delta(k - k')$. 
In addition, the modes which multiply these operators in Eq.~(\ref{fop}) are orthonormal: their
 norm, \eqr{KGnorm}, is $\delta(k-k')$, \emph{irrespectively} of both whether $v$ is sub- or supersonic,
and the choice of the dispersion relation (\ref{reldisp}).\footnote{In the case of subluminal dispersion,
there could exist a maximal value of $k$ associated with the vanishing of $\Omega^2(k)$.
In this case, the mode basis is no longer complete in a Fourier sense. 
To deal with this, one can either work~\cite{Macher:2008yq} 
with ``regularized'' dispersion relations which do not develop imaginary values of $\Omega$,
or with a ``Bloch waves'' perspective~\cite{Jacobson:1999ay} where $\Omega$ is periodic in $k$-space.
}

The completeness of the mode basis is needed to verify that 
the equal time commutator 
\be
[\hat \phi(t,x), \hat \pi(t,x')] = i \delta(x-x'),
\label{ETC} \ee
is satisfied when $\hat \phi$ is given in \eqr{fop}
and where $\hat \pi = (\partial_t - v \partial_x)\hat \phi$. 
In the present representation, the proof is easily obtained by making use of the completeness 
(in the sense of Fourier analysis) of the exponentials $e^{ikx}$,
with $k$ real from $[-\infty, \infty]$. It should be stressed that the proof applies 
to all dispersion relations and to all (constant) values of $v$, both sub- and supersonic. 

When considering nonhomogeneous stationary flows 
one must use the conserved frequency $\om$ 
in place of the wave vector $k$ to label modes and operators.
So, in preparation for this case, let us study the change
of representation from $k$ to $\om$ in homogeneous flows.

\subsubsection{Homogeneous metrics, the $\om$-representation}

The above conclusion concerning the completeness character of the modes 
 implies that, in any homogeneous flow, 
the ``extra'' (growing and decaying) solutions of 
the mode equation at fixed $\om$
must be discarded when quantizing $\hat \phi$. 
When the flow is subsonic, using the dispersion relation (\ref{reldisp}), 
one should thus discard the $2p$ growing/decaying modes, 
whereas when the flow is supersonic, one should discard only $2(p-1)$ modes.
Because of this, the change of representation from $k$ to $\om$ 
should be done separately in sub- and supersonic regions.

Let us first consider the 
subsonic case where only two real roots exist. 
For each $\om > 0 $, one 
has two modes:
\be
\varphi^i_{\om}(x) = \sqrt{\frac{dk^i}{d \om}}\,  \frac{e^{i k^i x}}{\sqrt{4 \pi  \Omega(k^i)}},\label{normmodes}
\ee
where $k^i= k^i(\om)$, $i=u,v$, are the two real roots of Eq.~(\ref{reldisp}) describing the right- and left-moving modes. These modes obey \eqr{waveeq} and are orthonormal for the KG product, in the sense of a Dirac distribution $\delta(\om - \om')$. 
Correspondingly, one has the rescaled annihilation operators:
\be
\hat a^i_\om = \sqrt{\frac{dk^i}{d \om} } \, 
\hat a_{k^i} ,
\ee
which verify $[\hat a^{i}_\om, \hat a^{j\dagger}_{\om'}] = \delta^{ij}\, \delta(\om - \om')$.
Using these modes, the integral in \eqr{fop}
can be rewritten as
\ba
\hat \phi(t,x) &=& \int_0^\infty d\om \left[ e^{-i\om t}\hat\phi_\om + h.c.\right],\label{sumoverom}\\
\hat\phi_\om &=& \varphi^u_{\om}(x)\,\hat a^u_\om + \varphi^v_{\om}(x)\, \hat a^v_\om.\label{just2}
\ea
The family of modes \eqref{normmodes} is complete,  because the exponentials $e^{i kx}$ are.

When the flow is supersonic, the situation is trickier
because the jacobian $dk/d\om$ crosses zero for right movers.
Nevertheless, similar results are obtained.
Namely, as above, one first separates the integral $\int_{-\infty}^\infty dk$ in \eqr{fop} into two integrals over 
a $u$ and a $v$ sector. When considering the $v$ sector ($k < 0$) 
in left-moving flows $v < 0$, nothing changes 
because $dk /d\om$ does not cross zero. Therefore, all left-moving (positive norm) modes can still be 
uniquely described by 
$\om$ belonging to $[0, \infty]$, 
and the change of variable $k\to \om$ can be made without further precaution.

The same is no longer true for the right-moving sector. 
In fact, the integral over the right movers $\int_0^\infty dk$ splits into an integral 
over $\om$ belonging to $[0, \infty]$ plus
another piece over negative frequencies belonging to 
$[-\om_{\rm max}, 0]$.
In addition, for a given value of $\om < 0$ in this
interval, two real roots $k^u(\om) > 0$ exist. 
It ends when the two roots merge
into each other when $-\om_{\rm max}$ is reached. 

Thus, for $ \om > \om_{\rm max}$, one 
has only one (positive norm) $u$-root and $\hat \phi_\om$
reads as in \eqr{just2}.
Instead, when $0 < \om < \om_{\rm max}$, 
three real roots exist: the continuation 
(in $\om$)
 of the former positive norm one, and 
two new roots with 
negative $\Om$, and thus negative norm, see Fig.~\ref{fig::merging}. 
In this case, $\hat \phi_\om$ must be decomposed as 
\be
\hat \phi_\om = \hat a^u_\om \, \varphi^u_{\om} + 
\hat a^v_\om \, \varphi^v_{\om} + 
\sum_{l=1,2} \hat a^{u \, \dagger}_{-\om, l}\,  \left(\varphi^u_{-\om, l}\right)^*.
\label{modeinomhom}
\ee
When compared with \eqr{just2}, the last two terms describe the two new roots.
A complex conjugate and a subscript $-\om$ have been used to characterize the
modes which multiply the creation operators $\hat a^\dagger_{-\om, l}$. 
This means that the modes  $\varphi^u_{-\om, l}$ have a 
positive norm and obey the mode equation with a frequency $i \partial_t = -\om < 0$.

Using the Fourier operator $\hat \phi_\om(x)$ of \eqr{modeinomhom}, 
containing both annihilation and creation sectors, 
the field $\hat \phi(t,x)$ can still be written as in \eqr{sumoverom}, as an integral
over $\om\in \, [0, \infty]$.

\subsubsection{Inhomogeneous metrics}

In metrics 
which contain a transition from a subsonic to a supersonic flow,
the decomposition of $\hat \phi_\om$ is modified: 
because of the scattering on $v(x)$, 
the asymptotic modes used above mix into each other, and with the growing and decaying modes.
As explained in Appendix~\ref{roledecay}, when 
excluding the  growing modes,  there is a 
change in the dimensionality of the mode bases. 
Indeed, instead of \eqr{modeinomhom}, for $0 < \om <\om_{\rm max}$, $\hat \phi_\om$ now reads
\be
\hat \phi_\om(x) = \hat a^u_\om \, \varphi^u_{\om}(x) + 
\hat a^v_\om \, \varphi^v_{\om}(x) + 
 \hat a^{u \, \dagger}_{-\om}\,  \left(\varphi^u_{-\om}(x)\right)^*.
\label{modeinom}
\ee
We have introduced only one extra mode, and not two, 
because the combination of the former two $\varphi^u_{-\om, l}$ 
that is orthogonal to the 
above $\varphi^u_{-\om}$ in the asymptotic supersonic region
is unbounded in the subsonic region.
This is very similar to what is found with the Airy function: only one combination of two oscillatory modes in the allowed region 
 remains bounded in the forbidden region.

For $\om > \om_{\rm max}$ instead, 
nothing changes and $\hat \phi_\om$ should 
still be decomposed 
as in \eqr{just2}
because one has two (positive norm) real roots for all $x$. 

When considering subluminal dispersion, the situation is similar but 
with several changes. For $0 < \om <\om_{\rm max}$, 
there are four real roots in the asymptotic \emph{subsonic} part of the flow, 
and the mode operator $\hat \phi_\om$ should be decomposed again as in Eq.~(\ref{modeinom}). 
The difference with superluminal dispersion is that the two additional roots
are associated with the first term $\hat a^u_\om \, \varphi^u_{\om}$ and no longer with
the third one.

\subsection[]{Bogoliubov transformation\label{enlargedbogo}}

For stationary flows and stationary states, the properties of the 
fluxes
are all encoded in a ($\om$-block-diagonal) 
Bogoliubov transformation. This relates
the in modes $\varphi^{in}_\om, \varphi^{in}_{- \om}$, 
which are associated with the destruction operators annihilating the
initial vacuum
state,  to the out modes $\varphi^{out}_\om, \varphi^{out}_{-\om}$ 
which characterize the asymptotic particle content 
of the fluxes.

To obtain the asymptotic temporal behavior of the modes from their
 asymptotic spatial behavior, 
it suffices (see e.g. section 1.3. in~\cite{Brout:1995rd}) 
to consider wave-packets centered about the 
value of $\om$ under examination. 
In what follows we shall not distinguish single frequency modes
from the corresponding broad wave-packet. 
Therefore when discussing the 
initial (or final) behavior of a mode, it should be understood as that
of the corresponding wave packet.

Given that the flow $v$ becomes asymptotically constant for large $|x|$,
the identification of in and out modes is free of ambiguity. 
The in (resp. out) modes are defined as the modes 
with positive comoving frequency $\Om$ in the asymptotic past (resp. future) 
and in the regions where the flow is homogeneous. 
They are separated into $u,in$ modes (resp. $u,out$ modes) 
with $d\Om/dk>0$ in the asymptotic past (resp. future),
and $v,in$ (resp. $v,out$) modes with $d\Om/dk<0$. These conditions fix uniquely the 
complete set of 3 in (resp. out) modes. In this work, the state of the field 
is taken to be the in-vacuum, annihilated by the $a^{in}_\om$ associated 
with the in-modes. Physically, it corresponds to a state devoid of particles 
propagating toward the horizon in the regions where $v$ is constant. 
Notice finally that this definition of the in vacuum does not coincide
with that of the ``Unruh-vacuum''~\cite{Unruh:1976db} which rests on the use
of relativistic fields, and of an affine null parameter on the past horizon.

\subsubsection{Simplified case}

To see what the new aspects brought in by dispersive effects are, 
let us briefly describe the Bogoliubov transformation
when assuming that the left-moving piece $\hat a_\om^v \phi^v_\om$ in Eqs.~(\ref{just2}, \ref{modeinom})
decouples from the right-moving sector. 
In this case it is sufficient to consider only one relation amongst $u$ modes, for instance
that with $\om > 0$
\be
\varphi^{u, in }_{\om} = \alpha_{\om} \, \varphi^{u, out}_{\om} + \beta_\om \,  \left(\varphi^{u, out}_{-\om}\right)^*.
\label{bog2}
\ee
Indeed, the norm of $\beta_\om$ is given by both 
overlaps
\ba
\vert \beta_\om \vert^2 &=&  \vert (\varphi^{u, out\, *}_{-\om}, \varphi^{u, in }_{\om}) \vert^2= 
\vert (\varphi^{u, out\, *}_{\om}, \varphi^{u, in }_{-\om}) \vert^2 . \quad\quad
\label{oldb}
\ea
This square 
fixes the mean occupation number of phonons (found in the in vacuum) 
for both positive and negative frequency: $\bar n_{\om} = \bar n_{-\om} = \vert \beta_\om \vert^2$.
In fact, since each produced pair contains one quantum of frequency $\om$
and its partner of frequency $-\om$, there is no need to differentiate
between these two occupation numbers. 
We emphasize this simple property because it will be lost
when the coupling to $v$-modes is no longer neglected.

A significant difference due to dispersion is that $\bar n_{\om} = 0$ for $\om > \om_{\rm max}$ because there is no bounded 
mode $\varphi^{u}_{-\om}$ for superluminal dispersion (no $\varphi^{u}_{\om}$ for subluminal dispersion).

\subsubsection{General case with $u-v$ mixing}

Taking into account the coupling between right and left-moving modes,
the above situation changes as follows.

For  $\om > \om_{\rm max}$, one still has $\bar n_\om = 0$.
However, the remaining two modes are scattered by the potential, and this is described by a single equation
\ba
\varphi^{u, in }_{\om} = T_{\om} \, \varphi^{u, out}_{\om} 
+ R_{\om} \,  \varphi^{v, out}_{\om}, 
\label{fakebogoin}
\ea
where $| T_{\om}|^2 +|R_{\om}|^2 = 1$.
One thus has an elastic scattering of $u$ and $v$ modes, without 
spontaneous pair creation.

For $\om < \om_{\rm max}$, unlike what was found in Eq.~(\ref{bog2}), three equations are now needed 
to characterize the Bogoliubov transformation between 
in and out modes
\ba
\varphi^{u, in }_{\om} &=& \alpha_{\om} \, \varphi^{u, out}_{\om} + \beta_{-\om} \,  \left(\varphi^{u, out}_{-\om}\right)^*
+ \tilde A_{\om} \,  \varphi^{v, out}_{\om},\nonumber
\\
\varphi^{v, in }_{\om} &=& \alpha^v_{\om} \, \varphi^{v, out}_{\om} + B_\om \,  \left(\varphi^{u, out}_{-\om}\right)^*
+ A_{\om}\,  \varphi^{u, out}_{\om},\nonumber
\\
\varphi^{u, in }_{-\om} &=& \alpha_{-\om} \, \varphi^{u, out}_{-\om} + \beta_{\om} \,  \left(\varphi^{u, out}_{\om}\right)^*
+ \tilde B_{\om} \left( \varphi^{v, out}_{\om}\right)^*  . \quad  \quad
\label{bogoin}
\ea
As usual, the coefficients are given by the KG overlap of the corresponding (normalized) in and out modes, 
e.g. $\beta_{\om}= - (\varphi^{u, out *}_{\om}, \varphi^{u, in }_{-\om})$. 
Using the mode orthonormality, 
their normalization %of the coefficients 
immediately follows, e.g. for the first equation 
one gets $\vert \alpha_{\om} \vert^2 - \vert \beta_{-\om} \vert^2 +  \vert A_{\om} \vert^2 = 1$. 
In these expressions, the minus signs come from complex conjugated, negative norm, modes.

When working in the in vacuum,
the occupation numbers of out quanta,
are respectively
\ba
\bar n_{\om}  &=& \vert \beta_{\om} \vert^2, \nonumber\\
\bar n_{\om}^v & = &    \vert \tilde B_{\om} \vert^2, \nonumber\\
\bar n_{-\om} &=& \vert \beta_{-\om} \vert^2 + \vert B_{\om} \vert^2
= \bar n_{\om} + \bar n_{\om}^v 
\label{threeoccn}. 
\ea
To obtain these expressions 
one needs to compute the $in-in$ expectation value of the corresponding
out occupation number operator.
The three expressions follow when decomposing 
the out
operators in terms of in ones, and using relations amongst  
Bogoliubov coefficients.

Several remarks should be made.
First, to get these expressions
we have treated  the quanta emitted to the right,
which correspond to the outgoing Hawking radiation and which are described by $\varphi^{u, out}_{\om}$,
on the same footing as those emitted to the left which are described by $\varphi^{v, out}_{\om}$ and 
$\varphi^{u, out}_{-\om}$. 
When $v$ is asymptotically constant on both sides, 
there is no reason to treat them differently. 

Secondly, because of the $u-v$ coupling,
the numbers of $u$ quanta in general differ:
$\bar n_{-\om} \neq \bar n_{\om}$. In fact the meaning of 
$\bar n_{-\om} = \bar n_{\om} +  \bar n_{\om}^v$ is clear. 
It tells us that two channels exist to spontaneously produce 
$u$-quanta of frequency $-\om$:
either through the usual channel where the partner is a Hawking quantum reaching $x = \infty$, 
or through the new channel
where the partner is a $v$ mode. 
When this $B$ channel is negligible, {\it i.e.} when $\vert B_{\om} \vert^2 \ll \vert \beta_{\om} \vert^2$, 
one recovers the former situation where $\bar n_{-\om} = \bar n_{\om}$.

The $B$ channel was first described in~\cite{Corley:1996ar}, where  
it was also claimed that ``this particle creation has absolutely nothing to do with black holes.'' 
We do not agree because, in our settings, there is no 
reason either to treat differently the production of $u-v$ pairs weighted by 
$B_\om$ from Hawking radiation,
{\it i.e.} the production of $u-u$ pairs weighted by $\beta_\om$. 
Indeed, both $\bar n_\om$ and $\bar n^v_\om$ are generically non zero %J5whenever
when (and only when) 
the negative frequency $u$-modes 
$\varphi^u_{-\om}$ exist.\footnote{It is appropriate to raise the question 
whether the existence of the negative frequency modes $\varphi^u_{-\om}$ 
implies that the background is similar to that of a ``black hole''. 
In this respect it should first be pointed out that 
the definition of a black hole is inherently ambiguous in the presence of dispersion.
Nevertheless, what is always true [in a metric as in \eqr{metric}] is that whenever $\varphi^u_{-\om}$ exists, 
its group velocity $d \Om/dk$ is smaller than $v$ in some region of space, 
as it is the case in the inside region of a black hole when using relativistic fields.}
More precisely, unless one deals with a 2D massless
relativistic field, the coefficients $B_\om$ do not vanish. In fact, 
one generically has $B_\om \neq 0$ 
even when using the dispersionless mode equation of phonons in a 
Eulerian fluid~\cite{Balbinot:2006ua,Carusotto:2008ep,MacherParentaniBEC}.

Thirdly, there is a linear scattering between $u$ and $v$ modes of positive
$\om$ 
which is characterized by the coefficients $\tilde A_\om$ and $A_\om$ for $\om <  \om_{\rm max}$ and by $R_\om, T_\om$ for $\om > \om_{\rm max}$.
This is reminiscent of the ``grey body factors'' which are found when considering four dimensional black holes.
However the latter are computed on one side only of the horizon, whereas here
the coefficients relate asymptotic modes defined on both sides.

Finally, we emphasize the fact that \eqr{bogoin} 
describes the scattering 
whenever a stationary background contains a single horizon surrounded by two asymptotic homogeneous regions.
It applies indeed to {\it all} dispersive mode equations, 
both for sub- and superluminal relations.  It is only when
the $u-v$ mixing identically vanishes that \eqr{bogoin}  reduces to \eqr{bog2}.  

\subsection{Wave-packet propagation}

\begin{figure}
\includegraphics{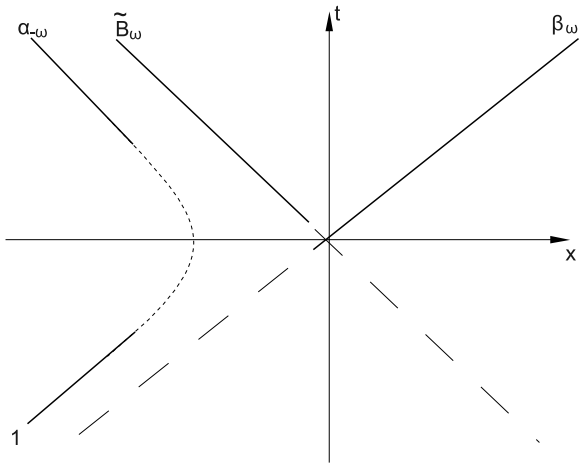}
\caption{\label{fig::uinmom}Schematic space-time behavior of a wavepacket made out of $\varphi^{u,in}_{-\om}$ modes, for superluminal dispersion.}
\end{figure}

It is useful to represent the above in 
modes. The description of out modes follows without difficulty.
Let us start with $\varphi^{u, in }_{-\om}$ in the superluminal case.
This mode has the following space-time stucture, 
see \figr{fig::uinmom}.
Initially, one only has the incoming branch which possesses a unit norm, and whose initial
wave vector is $k_+^u$, the largest additional real root when the flow is supersonic.
At late time, one has three branches: the ``reflected'' negative frequency mode described by $\varphi^{u, out }_{-\om}$
whose amplitude is $\alpha_{-\om}$ and whose wave vector is $k_-^u$, the smallest new root; 
the produced $v$ mode described by $(\varphi^{v, out }_{\om})^*$,
and the produced $u$ mode described by $(\varphi^{u, out }_{\om})^*$.
We leave the description of the other in modes and of the out modes up to the reader.

\begin{figure}
\includegraphics{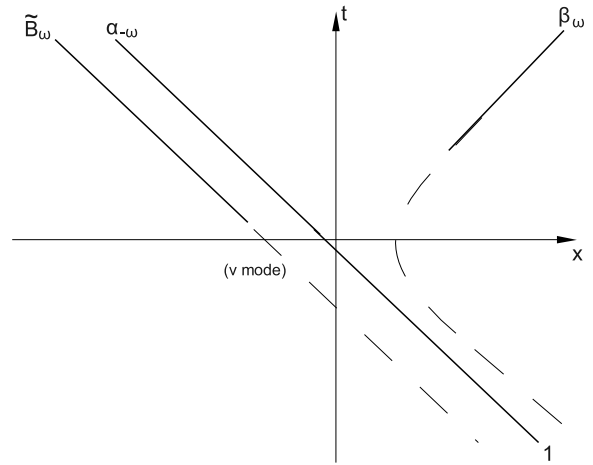}
\caption{\label{fig::uinmomsub}Space-time behavior of $\varphi^{u,in}_{-\om}$, for subluminal dispersion.}
\end{figure}

From the above we can already conclude that the emission of radiation stops for $\om > \om_{\rm max}$. 
Indeed, in the absence of negative frequency partners, both $\bar n_\om$ and $\bar n^v_\om$
identically vanish above $\om_{\rm max}$. The same is not true for the subluminal case.
It is thus worth describing the in mode $\varphi^{u, in }_{-\om}$ in this case,
see 
\figr{fig::uinmomsub}.
Initially, one still has only the incoming branch which possesses a unit norm
and which has the largest positive wave vector $k_+^u$.
At late time, one has again three branches:
the transmitted negative frequency mode described by $\varphi^{u, out }_{-\om}$
whose amplitude is $\alpha_{-\om}$, 
the produced $v$ mode described by $(\varphi^{v, out }_{\om})^*$, 
and the produced $u$ mode described by $(\varphi^{u, out }_{\om})^*$. 
When $\om > \om_{\rm max}$, the production of positive frequency $u$ modes vanishes ($\bar n_\om= 0$) 
but the `new' channel is still open. Thus, in the case of subluminal dispersion, one exactly has
$\bar n_{-\om} =  \bar n_\om^v$ for $\om > \om_{\rm max}$, and no longer 
$\bar n_\om= \bar n_{-\om} =  \bar n_\om^v= 0$ as for superluminal dispersion.

From a conceptual point of view, this completes the analysis of the black hole case.
What remains to be done is to compute the coefficients of Eqs.~(\ref{bogoin}).
If one is only interested in the flux in the in vacuum, the knowledge of the norm of $\beta_\om$ and $\tilde B_\om$ is sufficient
as they fix the three occupation numbers of Eq.~(\ref{threeoccn}).

Before performing their calculation, 
it is interesting to consider 
flows that engender white holes (WH). 
Given that $- v(x)$ describes a WH geometry when $v(x)$
described  a BH geometry,
the fluxes emitted by a WH 
can be algebraically related to those of the corresponding BH.

\subsection{White holes}

We first notice 
that the mode equation (\ref{weq0}) is left unchanged under the double replacement
$v \to -v$ and 
$t \to - t $. 
Then, in Fourier transform, the left-moving positive norm modes with fixed $\omega$ in the WH geometry are the complex conjugate of the right movers found in the corresponding BH geometry. This means that each mode has a group velocity $v_g = d\omega/dk$ opposite that of the corresponding BH mode, so that the BH in modes become out modes and vice versa. 

When working in the in vacuum (defined in the {\it same} way as in the former subsections),
and when
neglecting the coupling to the $v$-modes, this implies no change in the occupation 
numbers:
\be
\bar n^{\rm WH}_\om = \bar n^{\rm WH}_{-\om} = \bar n_\om, 
\ee
where $\bar n_\om$ is given by $\vert \beta_\om\vert^2$ obeying \eqr{oldb}.

Before discussing the physical consequences of this result, let us see how 
it gets modified when taking into account the $u-v$ mixing.
In this case
 the occupation numbers differ.
To obtain them, it suffices to 
write the inverse Bogoliubov transformation of \eqr{bogoin},
and relate the coefficients of the inverse transformation
to those of Eq.~(\ref{bogoin}). 
One finds that the occupation numbers 
in the WH geometry governed by $-v(x)$
are {\it exactly} given by 
\ba
\bar n^{\rm WH}_{\om}  &=& \vert \beta_{-\om} \vert^2, \nonumber\\
\bar n^{u\rm WH}_{\om} & = & \vert B_{\om} \vert^2,\nonumber \\
\bar n^{\rm WH}_{-\om} &=& \bar n_{-\om},
\label{threeoccnWH}
\ea
where the Bogoliubov coefficients are computed in the BH geometry described by $v(x)$ and where the superscript $u$ indicates that $|B_\om|^2$ corresponds to the occupation number of the right movers in the WH geometry.

The equality $\bar n^{\rm WH}_{-\om}= \bar n_{-\om}$ expresses that the (total) pair creation rate is
unchanged, as it must be the case 
because it governs the norm of the overlap between the
in and out vacua. The 
inequality $\bar n^{\rm WH}_{\om} \neq \bar n_{\om}$ means that their internal repartition does not coincide.
However, 
when $\vert \tilde B_{\om}\vert^2 \ll \vert \beta_{\om}\vert^2$,
one has $\bar n^{\rm WH}_{\om} \simeq \bar n^{\rm WH}_{-\om}= \bar n_{-\om}$. 
In this case, the WH fluxes are essentially the same as those of the corresponding BH fluxes.

Besides this close agreement, 
the important consequence of Eq.~(\ref{threeoccnWH})
is that dispersive theories predict that WH emit fluxes with well-defined asymptotic properties
(when starting with the asymptotic in vacuum, which is well-defined when
$v(x)$ is asymptotically constant). 
When using relativistic theories instead, one obtains an endless focusing of null 
geodesics toward the (past) horizon which prevents one from getting any outgoing 
radiation. 
(It should be nevertheless noticed that 
 in the equilibrium state, in the so called
Hartle-Hawking vacuum, the stress tensor of a relativistic field 
is regular and the spectrum of the
 particles ``emitted'' by the past horizon is thermal with the usual temperature,
in virtue of the stationarity.)

In spite of the near equality of occupation numbers in Eq.~(\ref{threeoccnWH}),
 there is a major physical difference between the WH and the BH cases,
which furthermore explains why dispersionless and dispersive theories behave so differently. 
In the WH geometry, 
the final values of the comoving (``proper'')
frequencies $\Om$ are 
of the order of the UV scale $\Lambda$, 
whereas, in the BH case, they 
are of the order of $\om \sim T_H$. 
This directly follows from the fact that 
the values of $\Omega$ and $k$
of WH quanta are those of the ancestors of Hawking quanta.
Because of this when taking the limit $\Lambda \to \infty$, {\it i.e.} the dispersionless limit,
the BH fluxes are asymptotically unchanged whereas the WH ones become
singular (and ill-defined).
This shows once more~\cite{Parentani:2002bd} that the relativistic (dispersionless)
case constitutes an isolated and unstable case, as far as UV properties are concerned.

In brief, we have shown that the 
ill-definedness of WH fluxes in relativistic theories, 
which is
due to the unbounded character of the blue shifting effect,
is properly regularized by the use of dispersion. 
(In this we disagree with the conclusion reached in~\cite{Leonhardt:2002yu}.)

\subsection{CJ modes}

We now address the question of the 
numerical analysis of the solutions of Eq.~(\ref{waveeq}). 
To this end,
one needs to introduce yet another type of mode. 
We shall call them the ``CJ'' modes,
since their usefulness 
for numerical analysis was first recognized in~\cite{Corley:1996ar}.
The basic reason to introduce these modes has to do with the 
control of the growing modes. Whereas conceptually we have demonstrated that the growing modes
should be discarded from the field operator, when performing a numerical integration
 of \eqr{waveeq}, they will be systematically generated and will thus completely mask the 
physical modes.
%R1 
To avoid this nuisance, one should use the CJ modes.
Even though they are defined essentially
in the same way for sub- and superluminal
dispersion, we analyze them separately because the information they carry differs.

\subsubsection{Subluminal dispersion}

\begin{figure}
\includegraphics{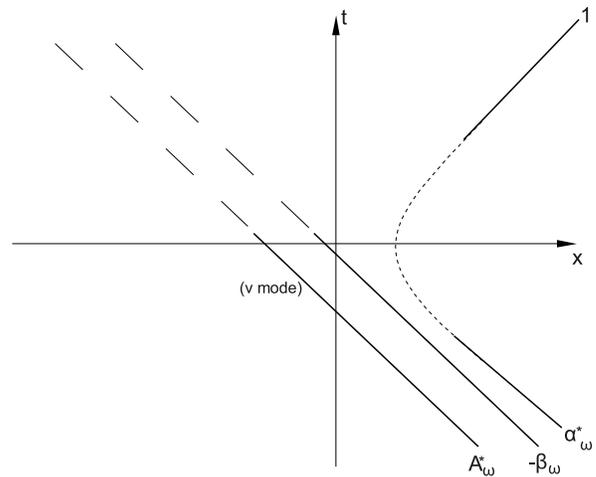}
\caption{\label{fig::uoutomsub}Space-time behavior of $\varphi^C_\om = \varphi^{u,out}_\om$, for subluminal dispersion.}
\end{figure}

Consider the integration from left to right
of \eqr{waveeq} starting deep in the supersonic region, {\it i.e.} on the left of the horizon,
where the complex-$k$ modes live. When reaching the horizon region, the solution 
is 
completely
dominated by the mode that grows when going toward the horizon (in our terminology, we called it
the decaying mode, see the Appendix). Therefore, one is effectively left with a well-defined mode, up to an overall normalization.
The CJ mode is the mode which 
has a unit positive KG norm, and such that only the coefficient of the decaying mode 
is non zero in the asymptotic region $x\to -\infty$. 
For $x \to \infty$ it contains four oscillatory asymptotic modes, see  \figr{fig::uoutomsub}.
It should be noticed that this is only true for quartic dispersion. Indeed for higher order
dispersion, see \eqr{reldisp}, there will still be $p-1$ growing modes which 
complicate obtaining a stable numerical analysis.

By direct inspection, one verifies that the CJ mode corresponds to the out mode $\varphi_{\om}^{u,out}$.
Using the conventions of Eqs.~(\ref{bogoin}), 
one has indeed
\be
\varphi^C_\om = \varphi_{\om}^{u,\, out} = \alpha_\om^* \, \varphi_{\om}^{u,\, in} - \beta_{\om} 
\left( \varphi_{-\om}^{u,\, in}\right)^* 
+ A^*_\om \, \varphi_{\om}^{v, in}.
\label{phiout}
\ee
Thus one can read off the occupation number
of Hawking quanta $\bar n_\om = \vert \beta_\om \vert^2$. 
However, using  the  CJ mode, 
one has no access to the number of $v$ quanta
$\bar n^v_\om = \vert \tilde B_\om \vert^2$. 
To have access to $\bar n^v_\om$ 
we shall use other modes, when the growth toward the horizon is not too strong.~\footnote{For high values
of $\om_{\rm max}/\kappa$ the growth of the CJ mode %R1
is so strong that we were not able to extract information
about the other modes. Instead, for 
$\om_{\rm max}/\kappa \lesssim 2$, the growth is sufficiently mild that 
we could compute accurately the two oscillatory solutions 
and thus extract the complete Bogoliubov transformation.}\setcounter{footref}{\thefootnote}

\subsubsection{Superluminal dispersion\label{CJsuper}}

\begin{figure}
\includegraphics[]{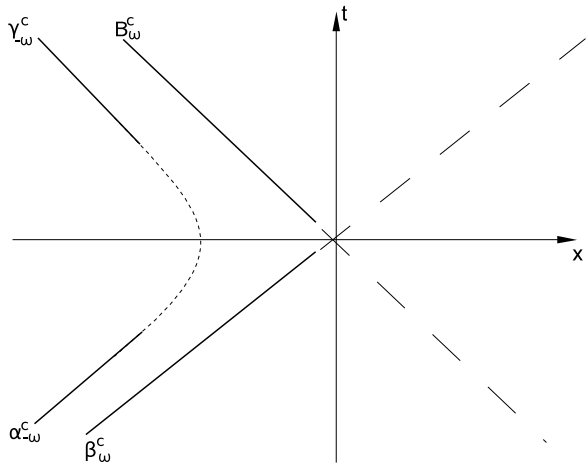}
\caption{\label{fig::corleysup}Space-time behavior of $\varphi^C_{-\om}$, for superluminal dispersion.}
\end{figure}

In the superluminal case, the situation is 
symmetrical with respect to the horizon, but also more complicated.
The difficulty arises from the fact that the 
 CJ mode does not correspond to an  in nor an out mode, 
since it possesses two incoming and two outgoing waves as shown 
in \figr{fig::corleysup}.
Nevertheless it is still defined 
as the unit positive norm solution that is purely decaying 
(growing toward the horizon) in the region where 
complex-$k$ modes live.
To have a positive norm, it must have a negative 
frequency. Hence
we shall denote it $\varphi^{C}_{-\om}$.
It possesses four branches for $x \to- \infty$ with well-defined asymptotic properties:
a wave packet made out of CJ modes has, at early times, the following behavior 
\be
\varphi^C_{-\om} = \alpha_{-\om}^C \, \varphi_{-\om}^{in} + \beta^C_\om \left(\varphi_{\om}^{u, in}\right)^*.
\ee
At late times, it behaves instead as
\be
\varphi^C_{-\om} = \gamma^C_{-\om} \, \varphi_{-\om}^{out} + B^C_\om \left(\varphi_{\om}^{v, out}\right)^*.
\label{phiCsup}
\ee
The coefficients obey 
$|\alpha_{-\om}^C|^2 - |\beta_\om^C|^2 = |\gamma_{-\om}^C|^2 - |B_\om^C|^2= 1$. 

We now have to establish the link between these coefficients and the occupation numbers
of out quanta in the in vacuum. 
Using the above equations, one can deduce 
\be
\bar n_\om =|\beta_\om|^2
 = |\beta^C_\om|^2 \times \left( 1 - |A_\om |^2 \right), \label{asuper}
\ee
and
\be
|B_\om|^2 = |B_\om^C|^2  \times \frac{|\alpha^v_\om |^2 }{1 + |B_\om^C|^2}. 
\ee
In the sequel we shall make the 
approximation (already used in~\cite{Corley:1997pr})
$\bar n_\om = |\beta^C_\om|^2$ to compute the
modifications induced by superluminal dispersion. 
This approximation shall be validated 
by showing that $|A_\om |^2 \ll 1$ is verified in the whole region of the parameter space where we could compute it, and that $|A_\om|^2 $ is decreasing toward 
the inaccessible region.

\section{Numerical procedure\label{numproc}}

\subsection{Initial conditions\label{initcond}}

As explained above, the presence of complex roots 
imposes the direction of integration and the way the initial conditions are fixed. Indeed, when integrating \eqr{waveeq} numerically, the mode growing in the direction of integration will dominate at some point the 
oscillatory modes. In the subluminal case, to get the CJ mode one must integrate from left to right starting from 
deep inside the supersonic
region.
For the same reason, to get the CJ mode in the superluminal case,
 the integration must be performed from right to left.

In addition, the imaginary part of $k^C_\om$ characterizing the 
CJ mode is generically large (in units of $\kappa$) 
even for moderate values of $D$ and $\lambda=\Lambda/\kappa$. 
For instance, for $D=0.5$, $\lambda=50$, and $\om=\kappa$, and for subluminal dispersion, 
$| {\rm Im} k^C| / \kappa \simeq 55$. 
This %R1 
constrains %1 on 
how far from the horizon the initial conditions can actually be set.

In practice, the initial conditions are thus fixed as follows. At a point $x_i$ deep into the supersonic (subsonic for superluminal dispersion) region, so that $v(x_i)$ is equal to its asymptotic value up to the machine precision ($10^{-16}$ with the $C$ double precision type), the value of the mode, % and 
its first, second and third derivatives are imposed using the fact that in this asymptotic region it is equal to 
\be
\varphi^C_{\om} = u_0 \, e^{ik^C_\om (x-x_i)}.
\ee
$u_0$ is 
taken very small so that the initial exponential growth of the mode does not cause an overflow 
(the $C$ double precision type is limited to the range $10^{-308}-10^{308}$). Since $u_0$ cannot be smaller than $10^{-308}$, $|x_i|$ is limited by 
the imaginary part of $k^C$. 
This limitation causes no problem since, as we saw, the bigger the growth of the CJ mode, 
the less important the precision of the initial conditions, 
the growth itself ensuring that any trace of unwanted excitation of the other modes quickly becomes smaller 
than the numerical noise. 

In brief we have two regimes, either $| {\rm Im} k^C| / \kappa \gg 1$ in which case the growth
is strong and we have only access to the CJ mode, or $| {\rm Im} k^C| / \kappa \lesssim 1$
and the growth is weak and we can compute all modes, see Footnote~\thefootref. 
Luckily, in the first case, the leading deviations wrt to the standard %J3fluxes
flux 
can be deduced from the CJ mode.

\subsection{Extraction of the asymptotic coefficients}
\begin{figure*}[!t]
\includegraphics{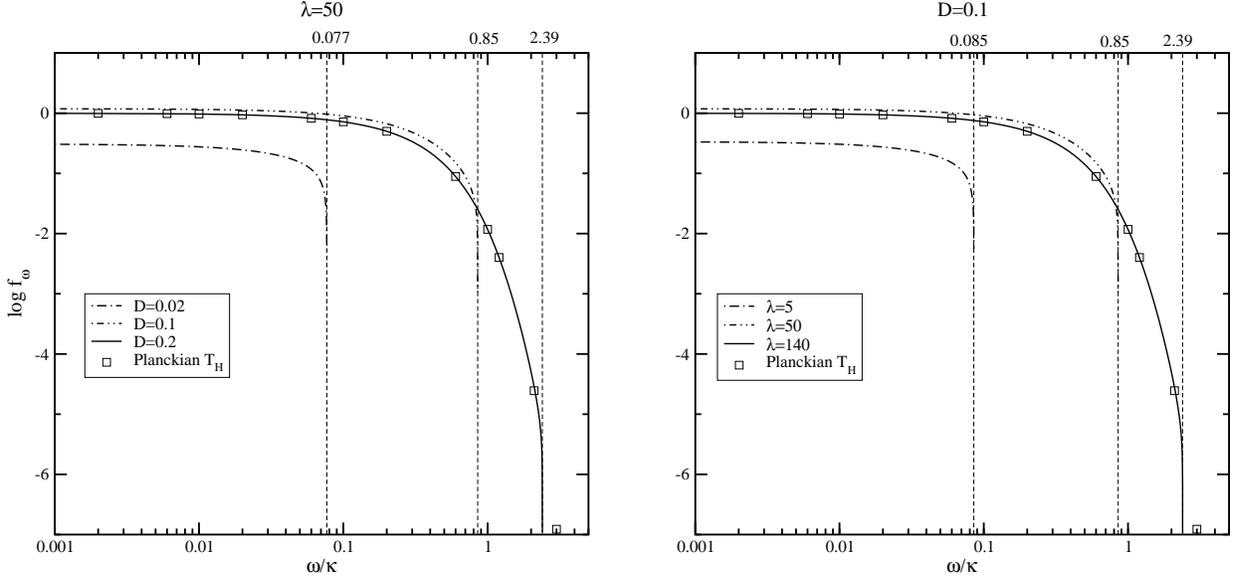}
\caption{Logarithm of the energy flux density received far from the horizon. The numbers at the top of the plots give the value of $\om_{\rm max}/\kappa$ for each curve. Left plot: $\lambda$ fixed to 50. $D=0.02, 0.1, 0.2$, from left to right. Right plot: $D$ fixed to $0.1$. From left to right, $\lambda=5, 50, 130$.
In both plots, the squares lie along the thermal energy flux with temperature $T_H = \kappa/2\pi$.\label{fig::fullspectrumF}}
\end{figure*}
The integration is carried out from $x_i$ to some $x_f$ on the other side of the horizon, by an embedded 8th order Runge-Kutta-Prince-Dormand algorithm, with a relative precision of $10^{-14}$. The real and imaginary part of the numerical solution to \eqr{waveeq} thus obtained are stored between some $x_w$ and $x_f$. $x_w$ is chosen so that it sits in the region where $v(x)$ has reached its asymptotic value.
It is typically taken equal to $-x_i$. $x_f$ is then chosen so that $|x_f-x_w|$ is equal to the period of the component with the smallest $|k|$, which on \figr{fig::asymptsols} is seen to be $k^v$,
the wave vector of the $v$-mode. There is thus no limitation  on the accessible domain of frequencies $\om$, besides the fact that for very low $\om$, one needs a lot of memory to store enough data to keep track of the component with the smallest wavelength, of order $\Lambda^{-1}$. For strongly dispersive cases, we could reliably access frequencies as low as $10^{-3} T_H$, where $T_H=\kappa/2\pi$ is the Hawking temperature.

On the interval $[x_w,x_f]$, the mode is a sum of the four asymptotic solutions:
\be
\varphi^C_\om(x) = \sum_{j=1}^4 c_j \, e^{i k_j x}.\label{expansion2}
\ee
Knowing the four roots $k_j(\om)$, 
we extract the coefficients $c_j$ by a least-square fitting procedure, tolerating relative errors on the various coefficients 
of less than $10^{-2}$. In practice the relative precision was usually much better, typically $10^{-5}$.
We are nevertheless limited by the precision of the numerical integration: the fit fails when the smallest coefficient is smaller than about $10^{-14}$ times the largest coefficient, which means that the smallest component in \eqr{expansion2} is at the level of the numerical error. This forbids one to approach arbitrarily close to $\om_{\rm max}$ since, as we shall see, the occupation number quickly drops when $\om \to \om_{\rm max}$. In the robust regime, this sets an absolute higher bound for the explorable values of $\om/\kappa$. Indeed in this regime, $\beta_\om/\alpha_\om \simeq e^{-2\pi\om/\kappa}$, so $\beta_\om=10^{-14}\alpha_\om$ 
is reached for $\om/\kappa\simeq 5$, {\it i.e.} $\om/T_H \simeq 30$.

The numerical solution to the wave equation is not normalized, 
and neither are the asymptotic solutions $e^{i k_j x}$. 
It does not matter since one only needs 
the relative norm of each component.
Taking into account the normalization of modes when $v$ is constant, see \eqr{normmodes},
we define 
\be
\eta^2 = |c_{u}|^2 \, \left(
\Om(k^u) \frac{d\om}{dk^u}
\right),
\ee
for the positive norm, low momentum $u$ component characterized by the root $k^u(\om)$.
In the subluminal case, the Bogoliubov coefficients are thus given by
\begin{align}
&|\beta_{\om}|^2  = \frac{|c^{u}_-|^2}{\eta^{2}} \, \,
\Om(k^{u}_-)
\frac{d\om}{dk_-} ,\label{betanum}\\
&|A_{\om}|^2  =\frac{ |c^{v}|^2}{\eta^{2}} \,
\Om(k^{v})
\frac{d\om}{dk^v}.
\label{Bnum}
\end{align}

The same equations hold for superluminal dispersion, 
with $\beta_{\om}$ replaced by 
$\beta^C_{\om}$, and $A_{\om}$ by $B^C_{\om}$, 
and with the normalization $\eta_{sup}$ given by
\ba
\eta^2_{sup} &=& |c^{u,out}_{-\om}|^2 \, \left(\Om(k^{u,out}_{-\om}) \frac{d\om}{dk^{u,out}_{-\om}}\right) -\nonumber\\ & &|c^{v,out}_{\om}|^2 \, \left(\Om(k^{v,out}_{\om}) \frac{d\om}{dk^{v,out}_{\om}}\right),
\ea
since the CJ mode is given by \eqr{phiCsup} at late times.

\section{\label{numressub}Results for subluminal dispersion}

\subsection{General properties of the spectra}

\subsubsection{Asymptotic energy flux}

The energy flux per $d\om$ and in units of $\kappa$, 
emitted to the right far from the horizon is equal to (denoting by $F$ the total energy flux)
\be
f_\om = \frac{4\pi^2}{\kappa}\frac{dF}{d\om} = \frac{\om}{T_H}|\beta_\om|^2 .
\label{energyflux}
\ee
The factor $4\pi^2$ is added for convenience, 
so that $f_\om \to 1$ for $\om \to 0$ in the standard 
case since $T_H = \kappa/2 \pi$.
This flux is represented in \figr{fig::fullspectrumF} for $\lambda=\Lambda/\kappa$ fixed to $50$, for values of $D$ from $0.02$ to $0.2$ (left plot), and for $D$ fixed to $0.1$ and $\lambda$ varying from $5$ to $130$ (right plot).

All curves exhibit a similar shape. When $\om\to 0$, 
$f_\om$
reaches an asymptotic constant value, 
that differs from $1$, the dispersionless value, 
but that becomes very close to it whenever $\om_{\rm max} \gg \kappa$.
When $\om_{\rm max}/\kappa$ is comparable to or smaller than $1$, 
the asymptotic energy flux can be 
either smaller or greater than the standard one. 
In addition, its variation with $D$ at fixed $\lambda$, or with $\lambda$ at fixed $D$, is not monotonic. When $\om$ approaches $\om_{\rm max}$, 
the energy flux quickly drops to zero, in agreement with the theoretical prediction that it vanishes
for $\om > \om_{\rm max}$.

\subsubsection{Mixing between right- and left-moving modes}
\begin{figure*}
\includegraphics{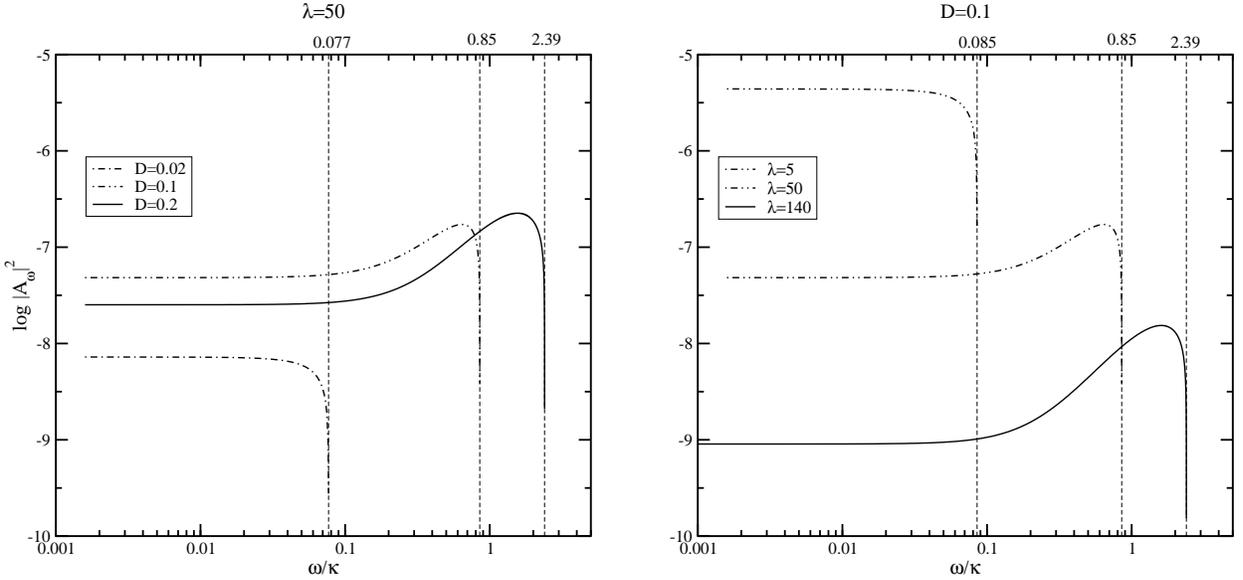}
\caption{Logarithm of $|A_\om|^2$ as a function of $\om$. In the left plot, $\lambda$ 
is fixed to 50, and in the right plot $D$ is fixed to $0.1$. The values of the parameters are identical to those in \figr{fig::fullspectrumF}\label{fig::fullspectrumA}}
\end{figure*}
\begin{figure}
\includegraphics{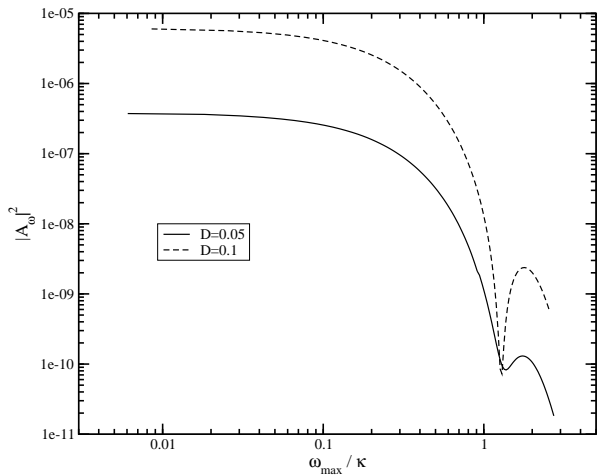}
\caption{$|A_\om|^2$ as a function of $\om_{\rm max}$, 
for $\om = 0.1 \times \om_{\rm max}$, 
for $D$ equal to $0.05$ (solid line) and 
$0.1$ (dashed line). 
The range in $\om_{\rm max}/\kappa$ shown 
corresponds to values of $\lambda$ between $0.5$ and $150$ for $D=0.1$, and between $1$ and $450$ for $D=0.05$.\label{fig::Ascaling}}
\end{figure}

In \figr{fig::fullspectrumA}, the norm $|A_\om|^2$
that measures the amount of elastic scattering between right-moving and left-moving modes, 
is represented as a function of $\om$. 
The parameters for the left and right plots are the same as in \figr{fig::fullspectrumF}. 
The curves 
display a characteristic shape, with a plateau at frequencies much smaller than $\om_{\rm max}$, a smooth increase as $\om$ nears $\om_{\rm max}$, that gets more pronounced as $\om_{\rm max}$ increases. Then, as for the energy flux, there is a sudden fall-off when reaching $\om_{\rm max}$. 

These properties can be qualitatively understood by considering the propagation backwards in time of the 
wave-packet of \figr{fig::uoutomsub}. As long as the evolution is adiabatic, there is no scattering 
into left-movers. The scattering must 
occur in the relatively well localized region of space near the turning point, 
where the WKB approximation breaks down. 
Its position depends on $\om$: when $\om\ll \om_{\rm max}$, it is very close to the horizon, and as $\om$ increases, 
it moves away from the horizon. When it is located in the region where 
$v(x)$ exits the linear regime with slope $\kappa$,
$\vert A_\om \vert $ reaches its maximal value. 
When $\om \to \om_{\rm max}$, the turning point 
enters into the flat region, and $A_\om$ goes to zero, as expected. 

In the right plot of \figr{fig::fullspectrumA}, 
the height of the low-frequency plateau 
significantly grows when $\lambda$ gets smaller. 
It is therefore interesting to further explore
the behavior of $|A_\om|^2$ for lower values of $\lambda$.
In \figr{fig::Ascaling}, this height is shown as a function of the cut-off 
frequency $\om_{\rm max}=\Lambda \, f_-(D)$. 
The growth seen in \figr{fig::fullspectrumA} saturates when 
$\om_{\rm max} \sim  0.1 \, \kappa$ 
for both values of $D$. 
We also notice that  $|A_\om|^2$ remains smaller than $10^{-5}$.

\subsubsection{Nonadiabatic effects}
\begin{figure}[!t]
\includegraphics{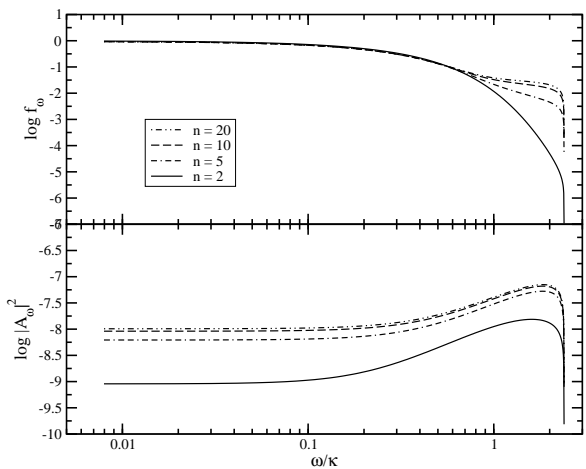}
\caption{Upper plot: logarithm of the energy flux density. Lower plot: 
logarithm of 
$|A_{\om}|^2$. $D=0.1$ and $\lambda=140$ in both plots.\label{fig::fullspectraN}}
\end{figure}
The power $n$ appearing in \eqr{vdparam} controls the sharpness of the transition between the linearly increasing flow velocity near the horizon and the constant velocity far from it. %R
It thus governs the adiabaticity of the wave propagation at the transition.
The qualitative reasoning above suggests that an increase of $n$ should lead to an enhancement of 
$|A_\om|^2$ and $f_\om$ short before $\om_{\rm max}$. \figr{fig::fullspectraN} shows that it is indeed the case.

From the upper plot in \figr{fig::fullspectraN}, 
we see that the modified  $f_\om$ contains a small contribution ($< 3 \%$ for $n < 20$) that shows 
up when the thermal, exponentially decreasing part of 
$f_\om$ is sufficiently suppressed~\cite{Corley:1996ar}. 
Moreover, we see that this contribution is directly related to
the nonadiabaticity since it increases with $n$. 
Instead, $\om \ll \kappa/2\pi$ 
the low-frequency part of the flux is only slightly affected, 
 as is the rapid fall-off for $\om \to \om_{\rm max}$.
For $|A_\om|^2$, a similar enhancement is observed, 
but it affects all values of $\om$.
This is not surprising, since $A_\om$ is entirely due to nonadiabatic effects.

\subsubsection{Detailed Analysis}

In the next sections we successively want to
\begin{itemize}
\item identify the region of the parameter space $(D,\lambda)$ where 
the outgoing flux $f_\om$ 
is robust, {\it i.e.}, where its value at $\om=T_H$,
differs little from the standard value
$=(e-1)^{-1}$.

\item determine, in the robust regime, 
the behavior of both the leading corrections to the thermal flux, 
and of the coefficient $|A_\om|^2$. 

\item analyze the spectral properties away from the robust regime. 

%R1
\item analyze $\bar n^v$ when the growth of the CJ mode is mild.

\item investigate the properties of the fall-off near $\om_{\rm max}$. 

\item analyze the integrated energy flux.

\end{itemize}

\subsection{Robust regime\label{robust}}

We define the robustness of the outgoing radiation by the fact that 
the energy flux, \eqr{energyflux}, differs little when evaluated around $\om = T_H$, 
from the thermal flux obtained without dispersion:
\be
f^{H}_\om = \frac{\om}{T_H}
\,\frac{1}{\exp(\om/T_H) - 1}\label{Planck}.
\ee
Our aim is twofold. First we wish to characterize the region of the parameter space where 
$f_\om$ of \eqr{energyflux}
hardly differs from $f^{H}_\om$.
Then we want to determine the scaling properties of the modifications.
 
To quantitatively study these aspects, we define $\Delta_H$ to be the relative difference 
between the modified and standard energy flux at $\om = T_H$: 
\be
\Delta_{H} = \left.\frac{f_\om-f^{H}_\om}{f^{H}_\om}\right|_{\om=T_H}. 
\label{DT}
\ee
It should be clear that the 
criterium $\Delta_H \ll 1 $ is local in that it is only concerned with what happens for frequencies near the Hawking temperature.
The smallness of $\Delta_H$ does not imply  
that $(f - f^H)/f^H$ remains small for all $\om$, as can be seen
from the upper plot in \figr{fig::fullspectraN}.

\subsubsection{$\om_{\rm max}/\kappa$ is the most relevant parameter}
\begin{figure}
\includegraphics{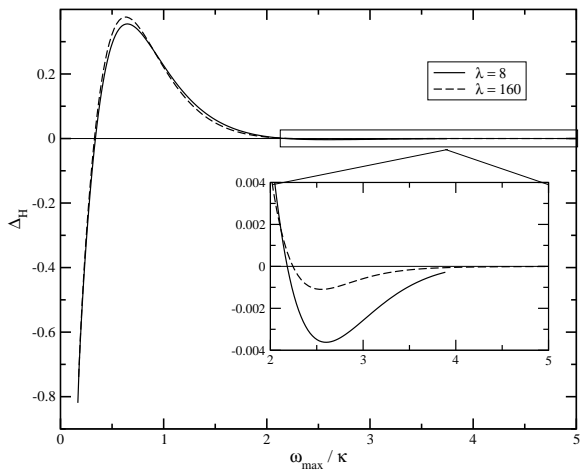}
\caption{$\Delta_H$ of 
\eqr{DT}, as a function of $\om_{\rm max}/\kappa$ for
 $\lambda = 8$ and $\lambda=160$. In spite of this factor $20$, 
the deviations are extremely similar for all values of $\om_{\rm max}$.
\label{fig::robust_criterium}}
\end{figure}
In \figr{fig::robust_criterium}, $\Delta_H$ is represented as a function of $\om_{\rm max}/\kappa$ for $\lambda=8$ and $\lambda=160$. Although 
the two values of $\lambda$ differ by a factor of $20$,
both curves stay
very close to each other even 
outside the robust regime. 
This confirms that $\Lambda/\kappa$ 
does not govern the robustness of the radiation. 
$\Lambda$ is nevertheless relevant 
since $\om_{\rm max}=\Lambda f(D)$. 
Hence $\Lambda/\kappa \gg 1$ is not a \emph{sufficient} condition for the robustness, 
but $\om_{\rm max}/\kappa \gg 1$ is.

With more details, starting from the right of \figr{fig::robust_criterium}
where $\om_{\rm max}/\kappa \gg 1$, we find as expected that $\Delta_H$ asymptotically 
vanishes. It stays much smaller than $1$ up to $\om_{\rm max}/\kappa$ of the order of 2. 
For smaller values of $\om_{\rm max}/\kappa$,  $\Delta_H$ becomes positive and reaches its maximal 
value, $\sim 25\%$,
for $\om_{\rm max}/\kappa \simeq 0.6$. 
Finally, when $\om_{\rm max}/\kappa \to 1/2\pi$ (and thus $\om_{\rm max}\to \om=T_H$), $\Delta_H \to 
-1$, which means that the flux vanishes, as already discussed.

\subsubsection{Leading corrections in the robust regime}

\begin{figure}
\includegraphics{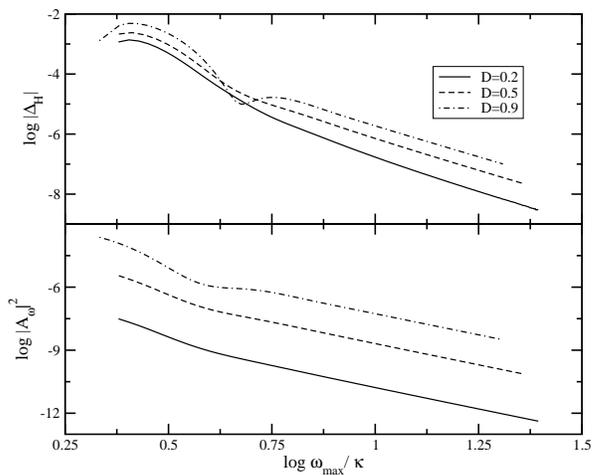}
\caption{$\log |\Delta_H|$ and $\log |A_{\om=T_H}|^2$ as a function of $\log \om_{\rm max}/\kappa$ for several values of $D$. When $\om_{\rm max} > 6$, both these quantities become linear, with a slope close to $-4$.\label{fig::robustdepDfixed}}
\end{figure}

To determine the scaling properties of the corrections, 
we have represented in \figr{fig::robustdepDfixed}, $\log |\Delta_H|$ 
as a function of $\log \om_{\rm max}/\kappa$ for several fixed values of $D$, 
That is, for each curve, $\lambda$ 
varies along the curve. 
The minimum value of $\om_{\rm max}$ is such that $\Delta_H$ is already in the region where it is small and negative
(see \figr{fig::robust_criterium}). 

We see that, whatever the value of $D$, for $\om_{\rm max}> 6\kappa$ 
($\log \om_{\rm max}/\kappa \gtrsim 0.75$), 
$\log |\Delta_H|$ is a monotonic linearly decreasing function of $\log \om_{\rm max}$. 
The fact that, for different values of $D$, the linear regime starts almost at the same 
$\om_{\rm max}$ is a further illustration of its 
relevance. Had we chosen $\log \lambda$ instead, 
the linear regime would start at very different abscissa for each value of $D$.

In the linear regime, the slope is of the order of $-4$, but slightly varies with $D$. 
The precise values of the slopes are $-4.42$, $-4.18$, $-4.10$ for $D=0.2,\, 0.5,\, 0.9$, respectively, so the slope gets closer to $-4$ for a higher $D$. 
Suprisingly, a similar behavior is observed for $\log |A_{\om=T_H}|^2$.
The slope is slightly less sensitive to $D$, as it is equal to $-4.05,\, -4.03,\, -4.01$ respectively. 

The fact that the robust regime, $\Delta_H \ll 1$, 
is already reached for $\om_{\rm max}/\kappa \geq 2$, 
together with the fact that the slope of $\log \Delta_H$ 
is of the order of $-4$ (and not $-1$ or $-2$ as one 
might have expected, and as was found for the power spectrum in inflationary cosmology~\cite{Macher:2008yq})
demonstrate
 in a precise and quantitative manner
that the thermal properties of Hawking radiation are hardly affected by $UV$ dispersion.

\subsubsection{Role of the asymptotic velocity $v_+ = -1 + D$.}

\begin{figure}
\includegraphics{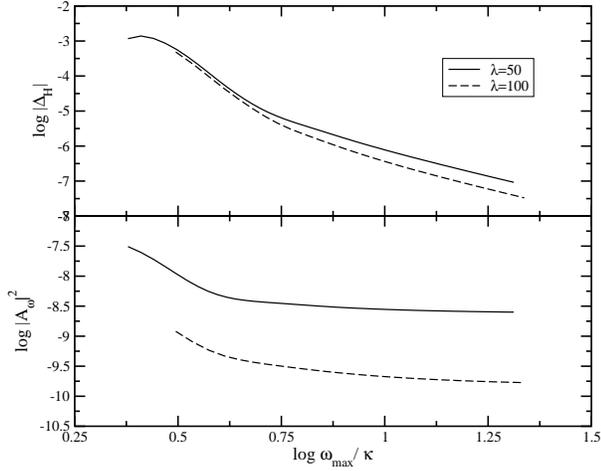}
\caption{$\log |\Delta_H|$ and $\log |A_{\om=T_H}|^2$ as a function of $\log \om_{\rm max}/\kappa$ for several values of $\lambda$. When $\om_{\rm max} > 6$, both these quantities become linear. $\log |\Delta_H|$ has a slope close to 3 for both values of $\lambda$, while the slope of $\log |A_{\om=T_H}|^2$ is very small and is $\simeq 0.14$ for $\lambda=50$ and $\simeq 0.29$ for $\lambda=100$.\label{fig::robustdepLfixed}}
\end{figure}

Figure~\ref{fig::robustdepLfixed} shows the same quantities, 
but now with a fixed value of $\lambda$ for each curve, while $D$ varies along the curves. 
Linear regimes starting at $\om_{\rm max}/\kappa \simeq 6$ are 
again obtained but the slope is of the order of $-3$ for $\log |\Delta_H|$, 
whereas it is only $-0.2$ for $\log |A_{\om=T_H}|^2$. 
(More precisely, the slope for $\log |\Delta_H|$ is $-2.95$ for $\lambda = 50$ and $-3.05$ for $\lambda=100$,
whereas, for $\log |A_{\om=T_H}|^2$, it is $-0.14$ and $-0.29$ respectively.) 
$D$ thus affects very differently the pair creation rate and the elastic scattering. 
This can be understood as follows. 
Since $D$ fixes the size of the near horizon region
where the gradient of $v$ can be approximated by a constant, 
$D$ governs the amount of redshift suffered by right movers from their turning point
to infinity. One thus expects that the deviations from thermality are highly sensitive to $D$, and this is
indeed the case since the log-log power is near to $-3$. 
On the contrary, the scattering between right and left movers is not related to the 
size of the near horizon region. 
Therefore one does not expect that 
$\log |A_{\om=T_H}|^2$ be strongly dependent on $\log D$. The numerical result $\sim -0.2$
confirms this expectation.

\subsubsection{Conclusion and comparison with former work}
\begin{figure}
\includegraphics{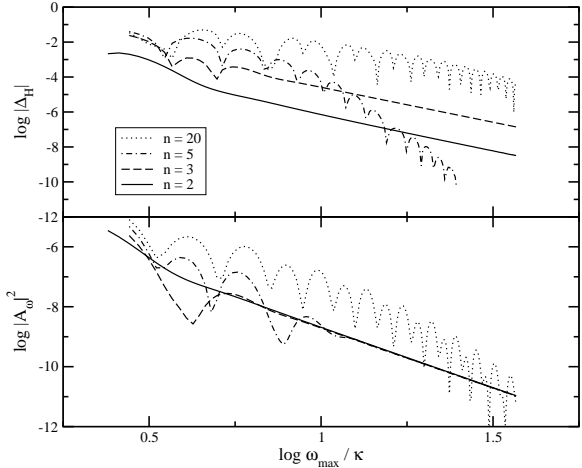}
\caption{$\log |\Delta_H|$ and $\log |A_{\om=T_H}|^2$ as a function of $\log \om_{\rm max}/\kappa$ for $D=0.5$ and different values of $n$.\label{fig::robustdepN}}
\end{figure}

Since the slopes obtained in \figr{fig::robustdepDfixed} and \figr{fig::robustdepLfixed} differ, $\Delta_H$ does not scale as a power of $\om_{\rm max}/\kappa$ alone. 
Remembering that $\om_{\rm max} = f(D)\Lambda$, 
we conclude that deep in the robust regime, the first corrections to the thermal spectrum can be written as:
\be
\Delta_H = - g(D, \Lambda/\kappa)
\times\left(\frac{\kappa}{\om_{\rm max}}\right)^{3}
\times\left(\frac{\kappa}{\Lambda}\right), 
\ee
where $g(D,\Lambda/\kappa)$ is positive and varies only slowly with the parameters. 

This result differs from 
what was reported in~\cite{Corley:1996ar}.
In that paper, $D$ was fixed to $0.5$, and the scaling of $\Delta_H$ with respect to $\lambda$  
was reported to be characterized by a power close to $1$. 
However there is no real contradiction as can be understood from
\figr{fig::robustdepN}, 
where the same quantities are represented as in the previous figures, for $D=0.5$, and several values of $n$. 
We see that, as $n$ increases, the superimposed oscillations become more 
pronounced and last until higher values of $\om_{\rm max}/\kappa$. 
In fact we see two regimes. When $n$ is 
smaller than about 4, the linear regime with a slope close to $-4$ remains.
Instead, for $n=5$ 
the nonadiabatic effects dominate and the linear regime is lost. In this sense, the low values of $n$ correspond to a {\it superrobust} regime, with well-defined first corrections to thermality
and with a clear scaling in the parameters. 
In~\cite{Corley:1996ar}, 
the scaling was estimated for a kinked velocity profile, which corresponds to the limit $n\to \infty$. 
In this limit, the nonadiabatic effects largely dominate and this 
probably
explains
why the authors missed 
the scaling as $(\lambda)^{-4}$. 

It is also interesting to note that the behavior of $\log |A_{\om=T_H}|^2$ is more robust to changes in $n$
since the linear regime exists longer, and the slope and intercept 
are not significantly modified.

\subsection{Black hole radiation outside the robust regime}
\label{smallom_nonrobust}

\begin{figure}
\includegraphics{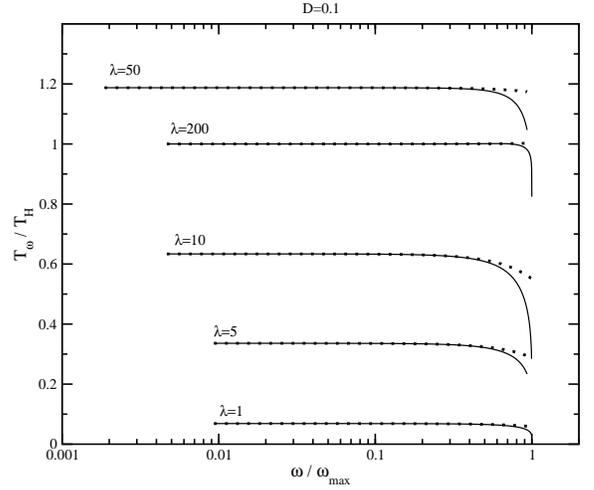}
\caption{\label{fig::TomD01}Effective temperature $T_\om$ as a function of $\om/\om_{\rm max}$, for $D=0.1$ and various values of $\lambda$. The dots show the fits obtained using the ansatz \eqr{Tomansatz}.}
\end{figure}

In the strongly dispersive regime, if one can expect large deviations, {\it i.e.} $O(1)$,
wrt to the standard flux of \eqr{Planck}, one has {\it a priori} no idea of what 
the properties of these deviations could be, nor how generic they are. 
Moreover, we do not think that they can be computed analytically. 
However with our code we were able to compute them 
when $\om_{\rm max}/\kappa \lesssim 1$, 
and, to our surprise, we found that thermality is preserved in the low
frequency part of the spectrum. 
In addition, because of the smallness of $\om_{\rm max}/\kappa$,
the growth of the CJ mode is moderate, and this allowed us to compute the observable $\bar n^v_\om$ which 
we could not have access to in the robust regime.

%R1
\subsubsection{Right-moving flux $\bar n_\om$}

To characterize the modified properties of the flux 
we  found {\it a posteriori} that it is convenient to %R define an
use the effective temperature $T_\om$ defined by:
\be
\bar n_\om = |\beta_\om|^2 = \frac{1}{\exp(\om/T_\om) - 1}\label{defTom}.
\ee

In \figr{fig::TomD01}, $T_\om/T_H$ is represented as a function of $\om/\om_{\rm max}$ for $D=0.1$ and 
different 
values of $\lambda$. Besides the expected quick drop of the power 
when  $\om \to \om_{\rm max}$, two 
remarkable features emerge: first, for all values of $\om_{\rm max}$,
 $T_\om$ is nearly constant for $\om <  \om_{\rm max}/10$. 
Secondly, the asymptotic temperature $T_0 = T_{\om\to 0}$ strongly differs
 from $T_H$ when $\om_{\rm max}/\kappa$ is small.

These two properties suggest studying two types of corrections: a (global) temperature shift, characterized by 
$T_0 - T_H$,
and a deviation from thermality, characterized by the running of $T_\om$ in $\om$. 
To sort out these two effects, we fit the curves with 
\be
T_\om/T_H = T_0/T_H + R\left(\frac{\om}{\om_{\rm max}}\right)^s.\label{Tomansatz}
\ee
The fits are shown as dotted lines in \figr{fig::TomD01}. They coincide perfectly with the curves at low frequencies and start to depart 
only around $\om = \om_{\rm max}/2$. The fit parameters are found to be:
\begin{center}
\begin{tabular}{|c|c|c|c|c|}
\hline
$\lambda$ &
$\om_{\rm max}/\kappa$& $T_0/T_H$ & $R$ & $s$\\
\hline
200 &
3.4 & 0.99987 & $3.0 %2.967
\cdot 10^{-3}$ & $2.05%1
$\\
140 &
2.4 &0.9995 & $-5.5 %451
\cdot 10^{-3}$ & $1.96%56
$\\
50 &
0.85 &1.187 & $-1.8 %03
\cdot 10^{-2}$ & $2.06%4
$\\
10 &
0.17 &0.6337 & $-8.3 %279
\cdot 10^{-2}$ & 2.01%2
\\
5 &
0.085 &0.3360 & $-4.8 %781
\cdot 10^{-2}$ & 2.01%3
\\
1 &
0.017 &0.06857 & $-1.0 %03
\cdot 10^{-2}$ & 2.01%2
\\
\hline
\end{tabular}
\end{center}
As expected, when $\om_{\rm max}/\kappa$ is large, $T_0$ becomes very close to $T_H$. 
In all cases, the power $s$ is close to $2$, and the coefficient $R$ is much smaller than 1. 

We are now in a position to determine 
to what extent the modified spectrum can be considered thermal.
To this end, let us compare the temperature  shift 
\be
\Delta_0 = \frac{T_H-T_0}{T_H},
\label{deltazero}
\ee
with the running of $T_\om$, that we characterize by 
\be
\Delta_{T_0} = \frac{T_0 - T_{\om=T_0}}{T_0}.
\label{deltaT}
\ee
($T_{\om=T_0}$ is the value given by the true curve, not the fit.)
We find:
\begin{center}
\begin{tabular}{|c|c|c|c|}
\hline
$\om_{\rm max}/\kappa$ & $\Delta_0$ & $\Delta_{T_0}$\\
\hline
%200 %& 0.047 
3.4 & $1.3\times 10^{-4}$ & %$ \simeq 5\cdot 10^{-6}$
 $<$ num. prec.\\
%140 %& 0.067 %656 
2.4 & $5.4 %39
\times 10^{-4}$ & $3.1 %02
\times 10^{-5}$\\
%50 %& 0.22 %14 
0.85 & -0.19 %74 
& $7.8 %32
\times 10^{-4}$\\
%10 %& 0.59 %09 
0.17 & 0.37 %662 
& 0.061 %099
\\
%5 %& 0.63 %265 
0.085 & 0.66 %40 
& 0.077 %672
\\
%1 %& 0.64 %393 
0.017 & 0.93 %14 
& 0.083 %02
\\
\hline
\end{tabular}
\end{center}
When $\om_{\rm max}/\kappa$ is large, these numbers make more precise the statement made before that we recover a Planck spectrum with the standard temperature, truncated when $\om \to \om_{\rm max}$. Indeed, both the temperature shift and the running in $\om$ are extremely small.

When $\om_{\rm max}/\kappa$ becomes small enough, 
as it is the case for values $\lambda<50$ in the table, 
the departure from the standard flux becomes significant. Nevertheless,
there is a clear ordering between the temperature shift and the running, the former being always 
much bigger than the latter. Thus, as a first approximation, 
the modified spectrum 
can be seen as a Planckian spectrum with a nonstandard temperature, and truncated for $\om \to \om_{\rm max}$,
and this even far away from the robust regime.

\begin{figure}
\includegraphics{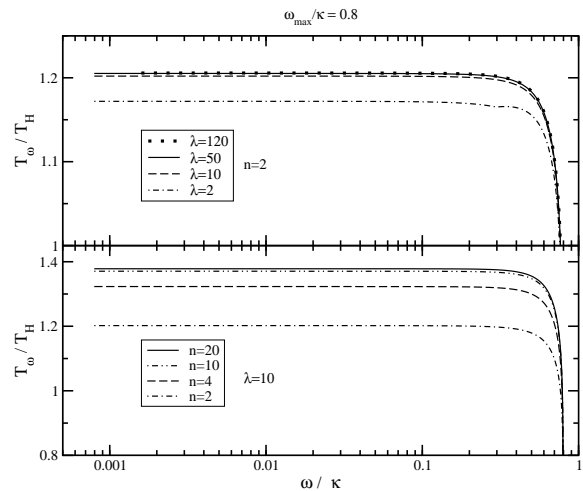}
\caption{\label{fig::Tomomx5}Effective temperature $T_\om$ as a function of $\om/\kappa$, for $\om_{\rm max}/\kappa=0.8$. Upper plot: $n$ fixed to 2, $\lambda$ varies; lower plot: $\lambda=10$, $n$ varies.}
\end{figure}

To complete our analysis, in \figr{fig::Tomomx5}, 
$T_\om$ is shown for several couples $(\lambda,D)$, 
such that $\om_{\rm max}/\kappa$ remain fixed to $0.8$. 
$T_0$ slightly varies ($\sim 3\%$) from one curve to another, and is thus not controlled 
only by $\om_{\rm max}$. However, the relative variations of $T_0$ are much smaller 
than the changes in $\lambda$. 
This shows what we had announced, namely that the modification of the spectra is essentially governed by $\om_{max}/\kappa$ and almost degenerate along the contours of constant $\om_{\rm max}/\kappa$ in the $(\lambda,D)$ plane.
In fact,
when $\lambda$ becomes high, $T_0$ saturates at a maximum value 
($\neq T_H$) which is truly fixed by $\om_{\rm max}/\kappa$ 
alone.
 
In the lower plot of \figr{fig::Tomomx5}, the effect of $n$ on $T_\om$ is investigated, for $\om_{\rm max}/\kappa=0.8$ and $\lambda=10$. The higher $n$, the higher $T_0$. The value of $T_0$ saturates at a maximum value when $n$ becomes high. We have also checked that the dependence of $T_0$ on $n$ is smaller when $\lambda$ is higher, as could be expected by looking at \figr{fig::fullspectraN}.

To conclude this subsection, we point out that the fact that
the temperature shift of \eqr{deltazero} is much larger than the running governed by $\Delta_{T_0}$
of \eqr{deltaT} for the entire class of metrics we considered 
severely limits 
the possibility, raised in~\cite{Schutzhold:2008tx}, 
that ``The $\om$-dependence of the Hawking temperature
can be explained by the fact that high-energy
wave-packets have a different group velocity than those
at low energy and hence the various modes `see' different
horizons and thus other values for the surface gravity''. 
Were this picture 
valid, the shift $\Delta_{0}$ would vanish since low-frequency wave-packets 
travel at the speed $c=1$ and thus see the standard surface gravity $\kappa=\partial_xv|_{x=0}$.

\subsubsection{Left-moving flux $\bar n^v_\om$\label{AddedNote}}

\begin{figure}
\includegraphics{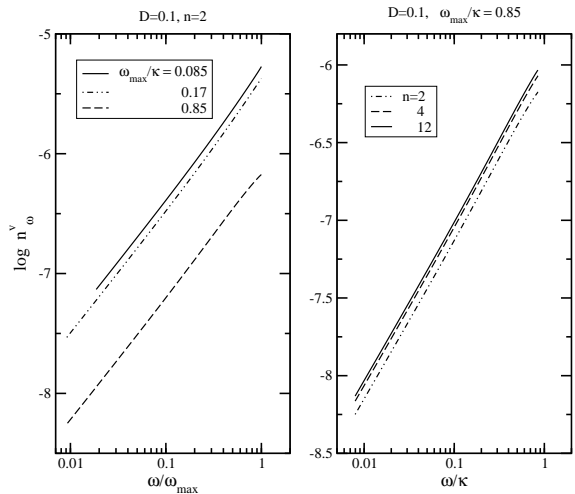}
\caption{Left plot: $\bar n^v_\om$ as a function of $\om/\om_{\rm max}$
for various values of $\om_{\rm max}/\kappa$, and with $D=0.1$ and $n=2$. 
Right plot: $\bar n^v_\om$ as a function of $\om/\kappa$
for various values of $n$, and with $\om_{\rm max}/\kappa= 0.85$, $D=0.1$.\label{fig::nvsub}}
\end{figure}

\begin{figure}
\includegraphics{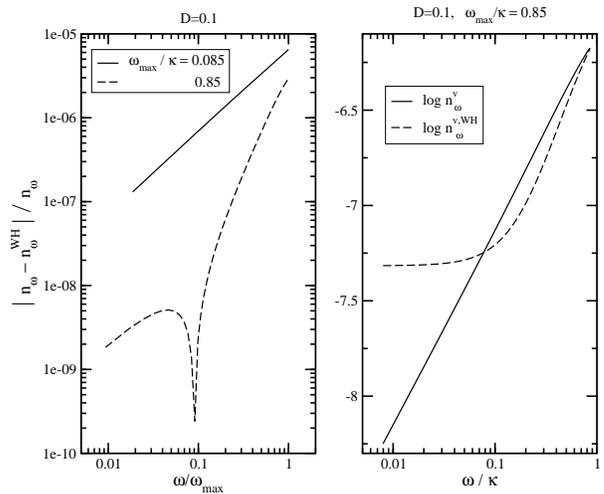}
\caption{Left plot: relative difference between $\bar n_\om^{WH}$ and $\bar n_\om$ as a function of $\om/\om_{\rm max}$, for  $\om_{\rm max}/\kappa = 0.085$ and $0.85$, and $D=0.1$. Right plot: $\bar n^v_\om$ and $\bar n^{v,WH}_\om$ as a function of $\om/\kappa$ for $\om_{\rm max}/\kappa = 0.85$ and $D=0.1$.\label{fig::WH}}
\end{figure}

When $\om_{\rm max}/\kappa \lesssim 1.5$, the CJ mode grows only slowly toward the horizon. %R1and 
It is thus possible to compute the other modes, % numerically,
 see Footnote~\thefootref. 
With three independent modes at hand, we can compute the %R1 all 
nine coefficients of the Bogoliubov transformation. Here, we present 
(for the %R1 very 
first time in the literature) 
quantitative results concerning the production of left-moving particles, governed by $\bar n^v_\om = |\tilde B_\om|^2$. It is represented in the left panel of \figr{fig::nvsub} as a function of $\om/\om_{\rm max}$
for $D$ fixed to $0.1$ and for 
%R1 two 
three
values of $\om_{\rm max}/\kappa$, namely $0.085$, $0.17$ and $0.85$. 
We note that $\bar n^v_\om$ is, to a good approximation, linear in $\om/\om_{\rm max}$. 
Hence it does not go to zero when $\om\to \om_{\rm max}$, in agreement with the fact 
that both $\varphi^{u,out}_{-\om}$ and $\varphi^{v,out}_{\om}$, 
and thus the coefficient $\tilde B_\om$ in \eqr{bogoin}, 
remain well-defined above $\om_{\rm max}$ in the subluminal case.

In the right panel of Figure~\ref{fig::nvsub} the influence of the nonadiabatic parameter $n$ %on $\bar n^v_\om$
 is investigated. As expected, the production of left-movers %depends %R1 on the adiabaticity and 
is more important when $n$ is higher. However the dependence on $n$ is weak and
%The effect of $n$ 
saturates when $n$ becomes high. 

\subsubsection{WH vs BH fluxes}

Having access to the full Bogoliubov transformation of \eqr{bogoin}, 
we can also compute the difference between the fluxes emitted by WH 
and BH. 
From the algebraic treatment, we can assert (using the unitarity of \eqr{bogoin}) 
that the difference of fluxes obeys
\be
 \bar n_\om^{WH} - \bar n_\om = - (\bar n^{v,WH}_\om - \bar n^v_\om).\label{nomminusnomWH}
\ee
However, to actually compute this difference, a numerical treatment is required.
The relative difference 
$(\bar n_\om^{WH} - \bar n_\om)/\bar n_\om$ 
is represented in \figr{fig::WH} for $D=0.1$ and 
$\om_{\rm max}/\kappa= 0.085$ and $0.85$. % 
This difference stays below $10^{-5}$. Moreover it %R1
decreases when $\om_{\rm max}/\kappa$ increases, {\it i.e.} when going toward the robust regime,
as one expected since it must vanish when the $u$-$v$ mixing itself vanishes.

For the larger value of $\om_{\max}/\kappa$ one can see that the relative difference changes sign.
This can be understood from \eqr{nomminusnomWH} and \figr{fig::nvsub}, where 
 we have seen that $\bar n^v_\om$ is proportional to $\om$ 
while $\bar n^{v,WH}_\om = |B_\om|^2$, shown in the right plot in \figr{fig::WH}, 
is constant at low frequencies. Thus one always has $\bar n_\om^{WH} > \bar n_\om$ at low enough frequencies. Notice also that contrary to the $u$ occupation numbers, 
$\bar n_\om^{WH}$ and $\bar n_\om$, 
neither of the $\bar n^v_\om$'s vanishes at $\om_{\rm max}$, but they become equal, ensuring that at all frequencies \eqr{nomminusnomWH} is satisfied.

Finally we notice that the relative difference
 does not diverge for $\om \to \om_{\rm max}$, which implies 
that $\bar n_\om^{WH}$ goes to zero near $\om_{\rm max}$ as quickly as $\bar n_\om$.

\subsection{\label{nearommax}Near $\om_{\rm max}$ region}

To complete our study of the spectrum, we investigate the quick fall-off near $\om_{\rm max}$. 
We find that, both for the energy flux and for $|A_\om|^2$, 
the fall-off presents a universal behavior (for the class of velocity profiles considered), independent of $\lambda$, $D$ and $n$.
Indeed we find that 
\be
f_\om = h_F(D,\lambda)\times \sqrt{\frac{\Delta\om}{\om_{\rm max}}},\label{ommfallF}
\ee
where $\Delta\om = \om_{\rm max} - \om$, and similarly for $|A_\om|^2$, 
with $h_F$ replaced by a different function of $D$ and $\lambda$, $h_A$.

\begin{figure}
\includegraphics{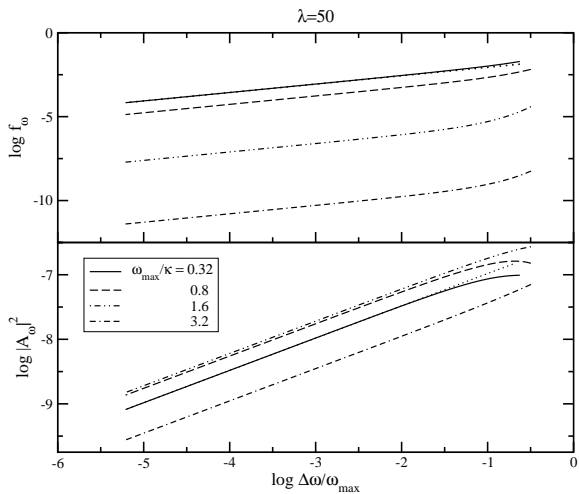}
\caption{$\log f_\om$ (upper plot) and $\log |A_\om|^2$ (lower plot) as functions of $\log \Delta \om/\om_{\rm max}$ for $\lambda=50$ and several values of $\om_{\rm max}/\kappa$. The legend applies to both plots. The dotted line has a slope exactly equal to 0.5 and fits perfectly the $\om_{\rm max}=2$ curves. All curves have the same slope in their linear region.\label{fig::nearomm}}
\end{figure}

\figr{fig::nearomm} shows $\log f_\om$ and $\log |A_\om|^2$ versus $\log \Delta \om/\om_{\rm max}$, near $\om=\om_{\rm max}$. $\lambda$ is fixed to 50 and each curve corresponds to a different value of $\om_{\rm max}$. Both quantities are linear when $\om$ is close enough to $\om_{\rm max}$, with a slope precisely equal to $0.5$, independently of the value of $\om_{\rm max}$. We verified that this behavior is also independent of $\lambda$. 
%%%
Finally we have also verified that the value of $n$, which strongly 
affects the behavior of the spectrum near $\om_{\rm max}$, as can be seen from \figr{fig::fullspectraN},
does not affect \eqr{ommfallF}.

\subsection{Integrated energy flux\label{integratedflux}}

\begin{figure}
\includegraphics{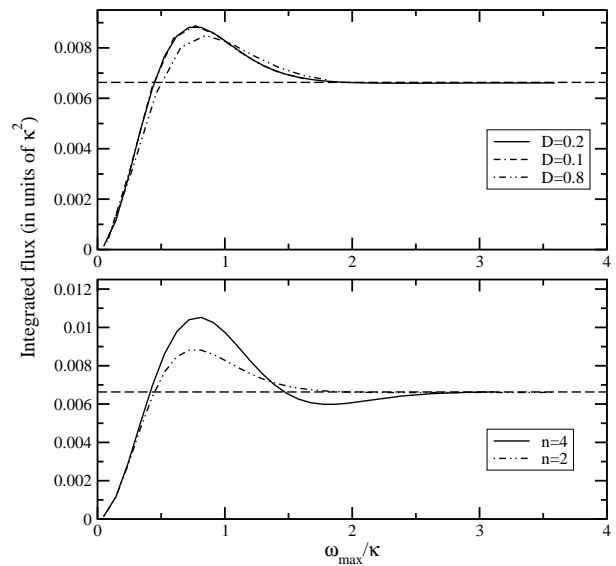}
\caption{Integrated flux in units of $\kappa^2$, as a function of $\om_{\rm max}/\kappa$. Upper plot: $n$ fixed to 2, several values of $D$. Lower plot: $D$ fixed to $0.2$, several values of $n$.
The horizontal line is the standard value ($= 1/48 \pi$) of the integrated flux obtained using the 
dispersionless theory.
 \label{fig::integflux}}
\end{figure}

In this section, we present the main features of the integrated energy flux 
obtained by integrating over $\om$ the differential flux $f_\om$.
It is computed as a Riemann sum 
and depends on $D$, $\lambda$, and $n$.
The sum is computed from $\om/\kappa = 10^{-3}$, where the asymptotic 
value of 
$f_\om$ is reached to a good precision, until $\om/\kappa = \om_{\rm max}/\kappa - 10^{-5}$. The step in $\om/\kappa$ is 
$0.01$, so that the relative variation between two successive points is small. 
Under these conditions, the Riemann sum gives a very good estimate of the integral.

The upper plot in \figr{fig::integflux} shows the integrated flux as a function of $\om_{\rm max}/\kappa$ for several values of $D$, and $n$ fixed to 2.
Like $T_{0}$ and $\Delta_H$, the integrated flux converges rapidly to its standard value for $\om_{\rm max}/\kappa>2$, 
even though the flux density $f_\om$
differs from the standard one for high frequencies. In other words, the nonlocal robustness criterium, that the integrated flux be close to standard, 
agrees with the local one defined in Sec.~\ref{robust}. This is because the exponential tail of the standard $f_\om$ contributes only marginally to the integral, so that the cut-off at $\om_{\rm max}$ in the modified flux does not cause noticeable differences in the integrated flux.

The lower plot of \figr{fig::integflux} shows that when $n$ increases, the amplitude of the oscillations in the integrated flux is more important, 
and 
they last until higher values of $\om_{\rm max}$. This reflects the fact that the nonthermal part of $f_\om$ 
yields a nonnegligible contribution to the integrated flux, and can be understood looking at \figr{fig::fullspectraN} where the enhancement of the high frequency part of the spectrum through nonadiabatic effects was shown.

\section{Results for superluminal dispersion \label{numressuper}}

\subsection{General properties of the spectra}

\subsubsection{Smallness of $u$-$v$ mixing.}

\begin{figure}
\includegraphics{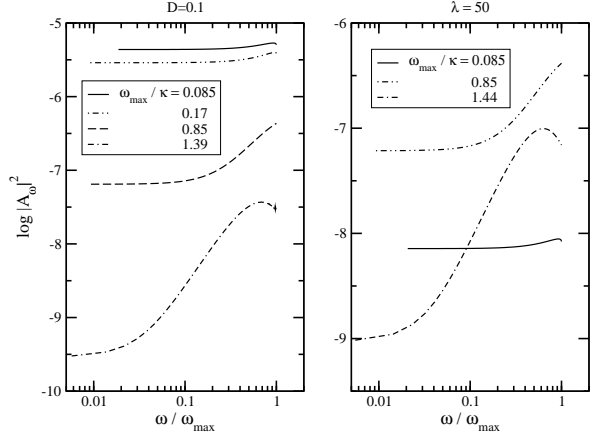}
\caption{The $u$-$v$ mixing coefficient $|A_\om|^2$
%R1
governing the difference between $|\beta^C_\om|^2$ and $|\beta_\om|^2$, see \eqr{asuper}, 
as a function of $\om$, for various values of $\om_{\rm max}$. Left plot: $D$ fixed to $0.1$. Right plot: $\lambda$ fixed to 50.\label{fig::fullspecsupA}}
\end{figure}

As can be seen from \eqr{asuper},
$|\beta^C_\om|^2$ is not exactly equal to $\bar n_\om$. 
In the following we make the approximation $\bar n_\om\simeq |\beta^C_\om|^2$, 
whose validity requires
 $|A_\om|^2\ll 1$. 
To justify this, 
we first show that the inequality is indeed verified everywhere %R1 where 
we can 
compute $|A_\om|^2$. 
$|A_\om|^2$ 
is shown in \figr{fig::fullspecsupA} for different values of $\om_{\rm max}$. 
At fixed $D$, on the left plot, it %R1
strongly decreases when $\om_{\rm max}/\kappa$ increases. 
For instance, it remains smaller than $10^{-7}$
for $\om_{\rm max}/\kappa= 1.39$. 
At fixed $\lambda$, on the right plot, 
it becomes monotonically decreasing only for high enough values of $D$, but stays always smaller than $10^{-6}$ for $\lambda=50$. 
These results %R1 
strongly 
suggest that it must be even smaller in the region of the parameter space where it is inaccessible to our code.
As we shall see in the next subsection,
the important result is that it is always much smaller than both %R1 
the modifications wrt the standard flux, and
 the relative difference between the super- and subluminal fluxes.
The approximation $|\beta^C_\om|^2\simeq \bar n_\om$ thus induces only a negligible error.

Note also that, unlike what was found in Figs.~\ref{fig::fullspectrumA} and~\ref{fig::fullspectraN},  
$|A_\om|^2$ does not vanish %R1 when approaching
for $\om \to \om_{\rm max}$ in the superluminal case. 
This is because $\varphi^{u,out}_\om$ and $\varphi^{v,out}_\om$ remain well-defined above 
$\om_{\rm max}$ so that, at $\om_{\rm max}$, $A_\om$ smoothly 
connects to $R_\om$ of \eqr{fakebogoin}. %R1

\subsubsection{Comparison between super-  and subluminal dispersion}

\begin{figure}
\includegraphics{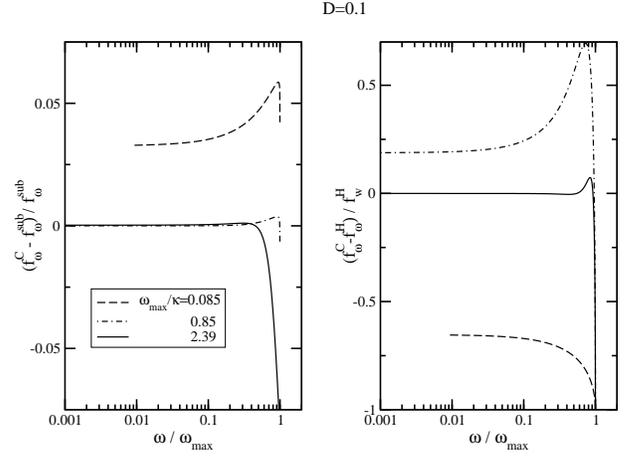}
\caption{Left plot: relative difference of the energy flux densities with superluminal and subluminal dispersion, as a function of $\om/\kappa$. The background geometry is fixed, with $D=0.1$ and $n=2$. $\om_{\rm max}/\kappa$ varies from 0.085 to 2.39. Right plot: relative difference between $f^C_\om$ and the standard energy flux $f^H_\om$, for the same set of parameters.\label{fig::subsupdiffD}}
\end{figure}

\begin{figure}
\includegraphics{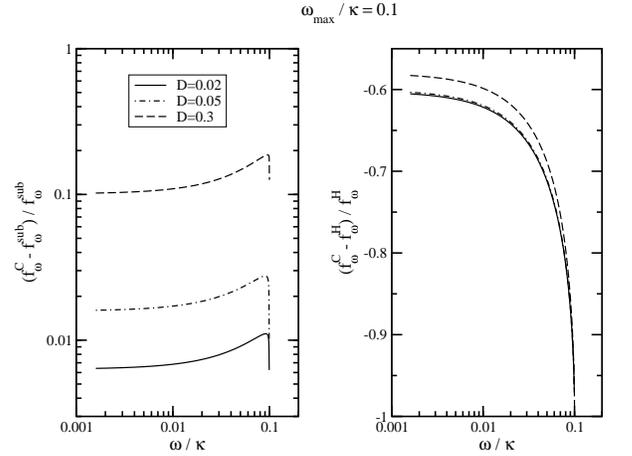}
\caption{Left plot: relative difference of the energy flux densities with superluminal and subluminal dispersion. $\om_{\rm max}/\kappa$ fixed to 0.1; $D$ varies from 0.02 to 0.3. Right plot: relative difference between $f^C_\om$ and the standard energy flux $f^H_\om$, for the same set of parameters.\label{fig::subsupdiffomm01}}
\end{figure}

In analogy with \eqr{energyflux}, we define the flux
\be
f^C_\om = \frac{2\pi\om}{\kappa} \, |\beta^C_\om|^2.
\ee
To our surprise, we found that 
$f^C_\om$ is extremely close to $f^{sub}_\om$ of \eqr{energyflux}
when working with the same background geometry ({\it i.e.} the same value of $D$ and $n$) and the same value of $\om_{\rm max}$. 
This unexpected similarity shows once more that 
$\om_{\rm max}$ is the relevant parameter for characterizing the departure from 
the dispersionless case.

Although $f_\om^{sub}$ of \eqr{energyflux} and $f_\om^C$ belong to two 
different models, it does make sense to study their difference when 
one works with the same velocity profile $v(x)$, so as to have the same behavior in the dispersionless regime, 
and the same value of $\om_{\rm max}$. (Under these conditions, the two values of $\Lambda$ cannot be exactly the same, 
but are close to one another.)
In \figr{fig::subsupdiffD}, the relative difference $(f^C_\om-f_\om^{sub})/f_\om^{sub}$ is represented versus $\om/\om_{\rm max}$ for a fixed $D=0.1$, and a series of increasing values of $\om_{\rm max}$.  The chosen values of $\om_{\rm max}/\kappa$ correspond to $\lambda=5, 50, 140$ respectively for subluminal dispersion, and slightly different values for superluminal dispersion, tuned to get the same $\om_{\rm max}/\kappa$.
The relative differences are of about $5\%$ when $\om_{\rm max}/\kappa=0.085$, 
but are extremely small for $\om_{\rm max}/\kappa=0.85$ and $\om_{\rm max}/\kappa=2.39$, %R1 but 
except in a thin region near $\om_{\rm max}$. 

The right plot in \figr{fig::subsupdiffD} also represents the relative difference between $f^C_\om$ and the standard energy flux. When the spectrum is not robust, as is the case for $\om_{\rm max}/\kappa<1$, the modifications with respect to the standard spectrum are much larger than the differences between the super- 
and subluminal spectra. In this sense, the modifications are controlled essentially by the value of $\om_{\rm max}/\kappa$, 
and seem to depend only marginally on the precise nature of the dispersion relation.

This point can be made more precise with the help of \figr{fig::subsupdiffomm01}, where the same quantities as in \figr{fig::subsupdiffD} are represented, but this time with $\om_{\rm max}/\kappa$ held fixed, equal to $0.1$, and a varying $D$. 
The values of $\lambda$ corresponding to $D=0.02, 0.05, 0.3$ are $64.84,\, 16.36,\, 1.09$ respectively, for subluminal dispersion (and of course slightly different for superluminal dispersion). From the right plot in that figure, it is clear that for this fixed value of $\om_{\rm max}/\kappa$, large variations of $(D,\lambda)$ cause only small variations in the corrections to the standard flux. In fact, as expected, the greater the scale separation between $\Lambda$ and $\kappa$, the more precise the statement that essentially only $\om_{\rm max}/\kappa$ matters and not the precise nature of the dispersion relation. This is illustrated by the left plot, where we see that as $D$ decreases (and thus as $\lambda$ increases to maintain $\om_{\rm max}/\kappa$ fixed), the relative difference between $f^C_\om$ and $f^{\rm sub}_\om$ becomes smaller and smaller.

The smallness of the differences in the energy flux between sub- and superluminal dispersion teaches us several lessons.
First, 
as soon as $\om_{\rm max}/\kappa \gtrsim 1$, the quantitative agreement, to a fraction of a percent, of the superluminal results with the subluminal ones (at least for frequencies not too close to $\om_{\rm max}$) means that the global properties for a given $\om_{\rm max}$ of the spectra for sub- and superluminal dispersion, such as the modified temperature, the running of the temperature, and the total energy flux, are the same to a very good approximation. Thus, we do not repeat in the case of superluminal dispersion the study performed in Sec.~\ref{smallom_nonrobust} and Sec.~\ref{integratedflux}, since the results 
are qualitatively and quantitatively the same.
Secondly, the robustness criterium and the 
range of robustness are the same for both types of dispersion.
That is, the radiation is robust for $\om_{\rm max}/\kappa>2$ in both cases.

\subsection{First corrections to HR in the robust regime}
\begin{figure}[!ht]
\includegraphics{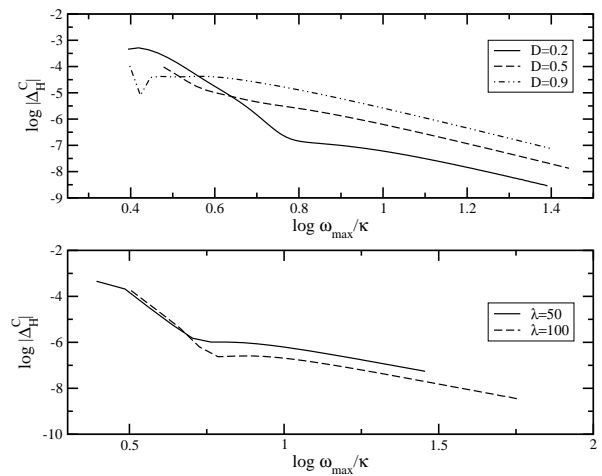}
\caption{Scaling of $\Delta^C_H$ with $\om_{\rm max}/\kappa$, for fixed $D$ (upper plot) and for fixed $\lambda$ (lower plot).\label{fig::robustdepSup}}
\end{figure}

The above agreement between $f_\om^{\rm sub}$ and $f_\om^C$ 
suggests that the scaling of the leading corrections with respect to the standard spectrum are also similar
for sub- and superluminal dispersion.
This point is investigated in \figr{fig::robustdepSup}, where the logarithm of
\be
\Delta^C_H = (f^C_\om - f_\om^H)/f_\om^H|_{\om=T_H}\label{DeltaC}
\ee
is represented versus the logarithm of $\om_{\rm max}/\kappa$ for several fixed values of $D$ and $\lambda$.

Like in the subluminal case, a ``superrobust'' regime is reached for $\om_{\rm max}/\kappa\gtrsim 6$. 
The slopes differ also slightly from the subluminal case. In the upper plot, we find $-3.46, -3.79, -3.88$ for $D=0.2$, $D=0.5$, $D=0.9$ respectively. In the lower plot, the slopes are $-2.42$ and $-2.48$ for $\lambda=50$ and $100$ respectively. 
The scaling in $\lambda$, at fixed $D$, of the first corrections is still with a power close to $-4$ in the superluminal case, while the scaling in $f_+(D)$ at fixed $\lambda$ is with a power between 2 and 3. It is also worth mentioning that $\Delta_H^C$ is negative in the superrobust regime, as was $\Delta_H^{sub}$. This is opposed to the first corrections to the inflationary spectra, that have different signs for sub- and superluminal dispersion (see~\cite{Macher:2008yq}).

\subsection{Production of left-movers}
\begin{figure}
\includegraphics{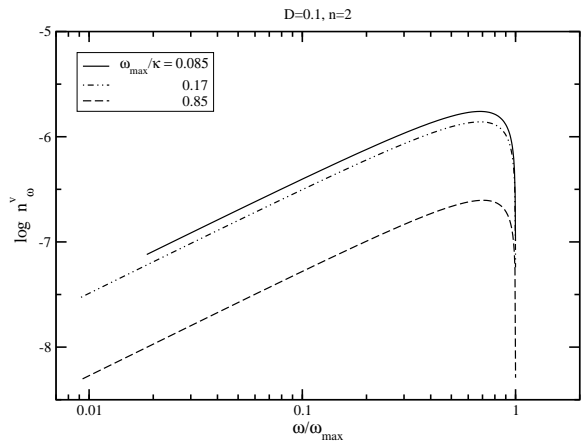}
\caption{$\bar n^v_\om$ as a function of $\om/\om_{\rm max}$, for  $\om_{\rm max}/\kappa$ from $0.085$ to $0.85$, with $D$ fixed to $0.1$.\label{fig::nvsup}}
\end{figure}

Figure~\ref{fig::nvsup} shows the occupation number $\bar n^v_\om$ for superluminal dispersion for $D=0.1$ and the same values of $\om_{\rm max}/\kappa$ as in \figr{fig::nvsub}. As in the subluminal case, it decreases when $\om_{\max}/\kappa$ increases. It is proportional to $\om$ for  
%when 
$\om\lesssim 0.1\times \om_{\rm max}$, 
but contrary to the subluminal case,
  it goes to zero when $\om$ approaches $\om_{\rm max}$. This is as expected
since the negative frequency partners of these $v$-quanta exist only for
$\om < \om_{\rm max}$.

\subsection{Near $\om_{\rm max}$ region.}

Finally, we have investigated the behavior of the fluxes near the maximum frequency. 
We find that both $n^v_\om$ and $f^C_\om$ decrease like $(\om_{\rm max}-\om)^{1/2}$ near $\om_{\rm max}$, as $f_\om^{sub}$ and $|A_\om^{sub}|^2$ did (see \eqr{ommfallF}).

\subsection{Comparison with Ref.~\cite{barcelo:024016}}

In Ref.~\cite{barcelo:024016}, when considering a time-dependent collapsing background metric, 
the authors reached the conclusion that Hawking radiation is no longer stationary when 
dealing with a quartic superluminal dispersion relation. 
We believe this conclusion is not due to the time-dependent character of the
geometry, but is rather due to the use
of a geometric optic approximation, in Eq.~(4.14),  
in a regime where it is not valid, as indicated by the 
numerical results of~\cite{Carusotto:2008ep}.
However a detailed calculation of the fluxes remains to be done.

\section{Conclusions\label{conclusions}}

We have studied the properties of the %R1 phonon 
fluxes emitted by 
acoustic black holes when taking into account the effects induced by
sub- and superluminal dispersion relations. We focused on one dimensional 
stationary flows which possess two asymptotic regions. In this case, unlike what is found
for gravitational black holes, the fluxes emitted on both sides of the 
horizon are well-defined observables. 

At the theoretical level, 
we showed that the dimensionality of the space of asymptotically bounded
 stationary modes is three, and that this space is complete.
This guarantees that the $3 \times 3$ Bogoliubov transformation of \eqr{bogoin} 
relating in and out modes is unitary. It reduces to a $2 \times  2$ matrix only if left and right-moving modes
completely decouple, which is a degenerate case, probably never found in condensed matter systems. 
We then showed that because of dispersion, 
there exists  a critical frequency $\om_{\rm max}$ above which
no radiation is emitted,
and the $3 \times 3$ transformation simplifies into \eqr{fakebogoin}. 
 This frequency depends both on the UV scale $\Lambda$
characterizing the dispersion, and on the asymptotic properties 
of the background geometry.
Because of the dispersion as well, the fluxes emitted by white holes
are well-defined and regular. Moreover we established how they are 
related to those of the corresponding black hole, see \eqr{threeoccnWH}
and \figr{fig::WH}. 
%R5

We have computed numerically the %R1 phonon 
fluxes in the case of quartic sub- and superluminal dispersion.
The main result of our analysis is the following. The deviations wrt the standard dispersionless fluxes
are highly degenerate along the lines of constant frequency $\om_{\rm max}$, see Figs.~\ref{fig::robust_criterium} and~\ref{fig::Tomomx5}, 
and, as a corollary, are essentially governed by  $\om_{\rm max}/\kappa$. 

Our second result concerns the characterization of the range of the robustness,
{\it i.e.} the range of the parameters 
within which the deviations are small.
We first found that for $\om_{\rm max} /\kappa > 6$, there is a superrobust regime
in which the leading deviations rapidly 
decrease as a power law in $\kappa/\om_{\rm max}$, 
where the power is equal to $3$ or $4$ depending on which parameter is held fixed, see Figs.~\ref{fig::robustdepLfixed} and~\ref{fig::robustdepDfixed}.
We also showed that for $\om_{\rm max} /\kappa > 2$, the deviations are already 
smaller than a percent. 

Our third important result concerns the absence of a significant 
running of the effective temperature, \eqr{defTom}, even %R1 
when leaving the robust regime. 
The low-frequency part of the fluxes, 
{\it i.e.} $\om < 0.1\times \om_{\rm max} $, is indeed well characterized by a 
common temperature that significantly differs, for $\om_{\rm max}/\kappa < 1$, 
from the standard temperature $\kappa/2 \pi$ 
but that hardly changes with $\om$,
 see the table after \eqr{defTom}.
This implies that the effective temperature $T_\om$
is probably not related in a universal manner 
to the gradient of the flow velocity $v$ evaluated at some $\om$-dependent horizon,
in agreement with the remarks made in~\cite{Corley:1996ar}.

Fourthly, we have also presented quantitative results concerning the production of $v$-quanta
in the domain $\om_{\rm max}/\kappa \lesssim 1$. 
We found numerically that 
$\bar n^v_\om$, is proportional to $\om$ at low frequencies, see Figs.~\ref{fig::nvsub} and~\ref{fig::nvsup}. 

Finally we showed in Sec.~\ref{numressuper} that the deviations obtained using sub- and superluminal dispersion
are unexpectedly similar when using theories having the same $\om_{\rm max}$. 
This means that the deviations wrt to the standard fluxes obtained with each theory 
are much larger than the difference between these deviations, typically by a factor of more than $10$,
see \figr{fig::subsupdiffomm01}.
However these conclusions are probably related to the fact that our velocity profiles
are symmetrical wrt the horizon, and possess two asymptotic regions. When considering 
profiles describing gravitational black holes, we expect that the deviations
due to  sub- and superluminal dispersion will differ more significantly. We also expect 
our results for subluminal dispersion to apply to gravitational black holes 
because the CJ modes essentially live in the outside region, as can be seen in \figr{fig::uoutomsub}.

\begin{acknowledgments}
We would like to thank the organizers and the participants 
of the workshop ``Towards the observation of Hawking radiation in condensed matter systems'' (http://www.uv.es/workshopEHR/) held at IFIC in Valencia in February 2009, where the extension of this work to Bose-Einstein condensates was presented, 
for many interesting remarks. We are grateful to Jihad Mourad
for discussions about the asymptotic bounded character of physical modes, and to Roberto Balbinot and Ted Jacobson for useful comments.
\end{acknowledgments}

\appendix
\section{Role of decaying modes\label{roledecay}}

In this appendix, our goals are, 
first, to show that in the black hole metric of  \eqr{metric}, 
when using the dispersion relation \eqr{reldisp} and for $\om<\om_{\rm max}$,
the space of spatially bounded modes has dimension three, 
and secondly, 
to understand the physical roles of the decaying modes.
In Sec.~\ref{modecompleteness}, when considering 
homogeneous flows, we saw that both growing and decaying modes
should be discarded from the field operator. 
When the flow is only asymptotically constant, 
as it is the case in \eqr{metric}, 
the symmetry between growing and decaying modes is broken 
because a mode that grows toward the horizon (we called this a `decaying' mode) 
does not necessarily grow in the other asymptotic region.

In the asymptotic supersonic region $x\to-\infty$ of \eqr{metric},
the wave equation reduces to the same form as in an homogeneous flow with velocity $v_-$. 
Thus, in this region, the general solution 
is a combination of all the exponential solutions that were present in the homogeneous case: four oscillatory solutions, $p-1$ decaying solutions, and $p-1$ growing solutions. This set 
defines the 
\emph{mathematically} complete basis of modes $\{b^{\ell,a}_\om(x)\}$ (thereafter called the ``left'' basis), where $a$ labels the roots of \eqr{reldisp} and $\ell$ stands for ``left'', such that each $b^{\ell,a}_\om$ has only one exponential component $e^{ik^a_\om x}$ in the asymptotic supersonic region. 
When restricting ourselves to spatially bounded modes, 
the $b^{\ell,a}_\om$ that correspond to growing components must be of course discarded. So one is left with $p+3$ independent modes bounded for $x\to-\infty$, numbered by $a=1. .p+3$, .

In the subsonic asymptotic region, in general, these modes are not bounded because they are 
 combinations of all exponential components, solutions of \eqr{reldisp} with velocity $v_+$.
On that side, one has $p$ growing and $p$ decaying modes, and only two oscillatory modes. 
In order to get modes that are everywhere bounded, one must consider combinations of the form
\be
\tilde \varphi^{j}_\om = \sum_{a=1}^{p+3} c^{j}_a \,  b^{\ell,a}_\om.
\ee
The condition that the 
coefficients of the growing components in the subsonic region be all zero yields $p$ constraints on the $p+3$ coefficients $c^j_a$. Therefore %R9 so that 
only three linearly independent bounded 
combinations exist. 
In other words, the dimensionality of the space of spatially bounded solutions is three for 
$0<\om<\om_{\rm max}$. This explains why only one extra-mode was added 
in \eqr{modeinom}. 
It is also important to note that the $p$ constraints in no way impose that the coefficients of the decaying modes be zero, and in fact the bounded modes, and in particular the ones constituting the in and out bases, generically contain decaying modes.

The physical role of these decaying modes is to `dress' the 
in and out modes in the  region %R9
 where $v(x)$
varies, 
in a way similar to the dressing of atoms 
by a local and nonpropagating polarization cloud when these are coupled to a radiation field~\cite{Massar:1993vg}.
Thus this dressing by decaying modes 
affects the properties of local observables (such as correlation functions) in the region where $v$ varies.

For completeness, a word on the construction of the in and out bases is in order. Just as we constructed the ``left'' basis, we can construct a ``right'' basis $\{b^{r,a}_\om\}$, from the asymptotic exponential solutions in the subsonic region. Looking at the space-time properties of wavepackets made out of the asymptotically oscillatory modes in both $\{b^{r,a}_\om\}$ and $\{b^{\ell,a}_\om\}$, one can reclassify them into a group of three in modes and one of three out modes (still unbounded at this point). Each of these two groups mixes ``left'' and ``right'' modes. The dimensionality three of the space of bounded modes then ensures that we can find bounded combinations with only one in (respectively out) mode with a nonzero coefficient. These combinations, suitably normalized and possibly complex conjugated to get a positive norm mode, yield the 
in (resp. out) basis used in the text. Note also that since there are only two oscillatory solutions in the subsonic region, there exists 
one (and only one, up to an overall factor) bounded mode that is purely decaying in this region. 
This mode, correctly normalized, is the  CJ mode, here generalized to arbitrary (polynomial) dispersion.

\bibliography{../biblio/biblio}

\end{document}